\documentclass[12pt,journal,onecolumn]{IEEEtran}
\usepackage[hang]{subfigure}
\usepackage{epsfig,rotating,setspace,latexsym,amsmath,epsf,amssymb,amsfonts,bm,theorem,epstopdf}
\usepackage{cite,authblk}
\usepackage{algorithm}
\usepackage[noend]{algpseudocode}
\usepackage{setspace}
\usepackage{color}
\usepackage{graphicx}
\usepackage{mathrsfs}
\ifCLASSOPTIONcompsoc
\usepackage[caption=false, font=normalsize, labelfont=sf, textfont=sf]{subfig}
\else
\usepackage[caption=false, font=footnotesize]{subfig}
\fi

\newtheorem{lemma}{Lemma}

\newenvironment{Proof}[1]{\medskip\par\noindent{\bf Proof:\,}\,#1}{{\mbox{\,$\blacksquare$}\par}}

\allowdisplaybreaks

\definecolor{green1}{rgb}{0.2,0.7,0.2}
\definecolor{brown}{rgb}{1,0.5,0.2}
\DeclareMathOperator*{\argmin}{\arg\!\min}

\newcommand{\defeq}{\ensuremath{\triangleq}}

\usepackage[colorinlistoftodos,textwidth=\marginparwidth]{todonotes}

\begin{document}
\title{Gradient Coding with Dynamic Clustering for Straggler-Tolerant Distributed Learning \thanks{This work was supported by EC H2020-MSCA-ITN-2015 project SCAVENGE under grant number 675891, by the European Research Council project BEACON under grant number 677854, and by CHIST-ERA grant CHIST-ERA-18-SDCDN-001 (funded by EPSRC-EP/T023600/1). This work was presented in part at the IEEE International Conference on Communications, Montreal, Canada, June 2021 \cite{Buyukates21a}.}}

\author[1]{Baturalp Buyukates}
\author[2]{Emre Ozfatura}
\author[1]{Sennur Ulukus}
\author[2]{Deniz G\"{u}nd\"{u}z}
\affil[1]{\normalsize Department of Electrical and Computer Engineering, University of Maryland, USA}  
\affil[2]{\normalsize Department of Electrical and Electronic Engineering, Imperial College London, UK}

\maketitle

\begin{abstract}

Distributed implementations are crucial in speeding up large scale machine learning applications. Distributed gradient descent (GD) is widely employed to parallelize the learning task by distributing the dataset across multiple workers. A significant performance bottleneck for the per-iteration completion time in distributed synchronous GD is \textit{straggling} workers. Coded distributed computation techniques have been introduced recently to mitigate stragglers and to speed up GD iterations by assigning redundant computations to workers. In this paper, we consider gradient coding (GC), and propose a novel dynamic GC scheme, which assigns redundant data to workers to acquire the flexibility to dynamically choose from among a set of possible codes depending on the past straggling behavior. In particular, we consider GC with clustering, and regulate the number of stragglers in each cluster by dynamically forming the clusters at each iteration; hence, the proposed scheme is called \textit{GC with dynamic clustering} (GC-DC). Under a time-correlated straggling behavior, GC-DC gains from adapting to the straggling behavior over time such that, at each iteration, GC-DC aims at distributing the stragglers across clusters as uniformly as possible based on the past straggler behavior. For both homogeneous and heterogeneous worker models, we numerically show that GC-DC provides significant improvements in the average per-iteration completion time without an increase in the communication load compared to the original GC scheme.
\end{abstract}

\section{Introduction}

Gradient descent (GD) methods are widely used in machine learning problems to optimize the model parameters in an iterative fashion. When the size of the training datasets and the complexity of the trained models are formidable, it is not feasible to train the model on a single machine within a reasonable time frame. To speed up GD iterations, gradient computations can be distributed across multiple workers. In a typical parameter server (PS) framework with synchronous GD iterations, the dataset is distributed across the workers, and each worker computes a gradient estimate, also called a \textit{partial gradient}, based on its own local dataset. The PS then aggregates these partial gradients to obtain the full gradient and update the model. In this distributed setting, the main performance bottleneck is the slowest \textit{straggling} workers. Many recent works have focused on developing straggler-tolerant distributed GD schemes. In these works, the main theme is to assign redundant computations to workers to overcome the potential delays caused by straggling workers, either together with coded dataset assignment to workers, i.e., coded computation \cite{CC.1,CC.2,CC.3,CC.4, CC.6, CC.7, Yu18, Dutta18c, CC.8, CC.11, CC.13, Ozfatura19b, Mallick19, Ozfatura18, Yang19, Yang19c, Park19, Bitar19, Sun19d, Hasircioglu20, Buyukates19c, Ozfatura20}, or combined with coded local computations, i.e., coded transmission \cite{UCCT.1,UCCT.2,UCCT.3, Zhang19, Kadhe19, Ozfatura19,Li19,Tauz19, Bitar20, Charalambides20, Raviv20}, or by simply using backup computations, i.e., uncoded computation \cite{UCUT.5, UCUT.2, UCUT.1, Ferdinand18b, BehrouziFar18, UCUT.4}. 

In this paper, we consider the gradient coding (GC) framework introduced in \cite{UCCT.1}, where the dataset is distributed across the workers in an uncoded but redundant manner, and workers return coded computations to the PS. We note that this can also model a scenario, in which data is collected directly by the workers, instead of being distributed by the server. Redundancy can either be created by data sharing among the workers, or may be inherent due to the data collection/generation mechanism. Thanks to the redundancy in the local datasets, partial gradients from only a subset of the workers will be sufficient to recover the full gradient. Coded combinations retrieved by the workers are designed such that any subset of responses from sufficiently many workers will allow the computation of the full gradient by the PS. Further details of GC are presented in the next section. 

To improve the performance of the GC scheme, reference \cite{Ozfatura19} proposes a static clustering technique, which entails dividing the workers into smaller clusters and applying the original GC scheme at the cluster level. This technique is shown to improve the average computation time compared to the original GC scheme. With clustering, unlike in the original GC scheme, the number of tolerated stragglers scales with the number of clusters when the stragglers are uniformly distributed among the clusters. However, this may not be the case in practical scenarios as evident in the measurements taken over Amazon EC2 clusters that indicate a time-correlated straggling behavior for the workers \cite{UCCT.1, Yang19}. In this case, the advantage of clustering diminishes since the stragglers are not uniform across clusters.

To mitigate this problem and to further improve the performance, in this paper, we introduce a novel GC scheme with dynamic clustering, called GC-DC. The main idea behind GC-DC is to assign more data samples to workers than the actual computation load (per-iteration) to give them the flexibility in choosing the computations they need to carry out at each iteration. This allows the master  to choose at each iteration which subset of computations each worker should try to complete, and which coded combination it should transmit back to the master. To reduce the potential solution space, we focus on the GC scheme with clustering, and let the master decide on the clusters to be formed at each iteration. At each iteration, GC-DC forms the clusters such that the stragglers are distributed across the clusters as uniformly as possible based on the workers' past straggling behavior. We numerically show that the proposed GC-DC scheme significantly improves the average per-iteration completion time without an increase in the communication load under both homogeneous and heterogeneous worker environments.

The rest of this paper is organized as follows: In Section~\ref{GC_cl}, we present the GC and GC with clustering frameworks. In Section~\ref{GC_dyn}, we introduce the GC with dynamic codeword assignment scheme to improve the average iteration completion time of the static GC schemes and present the problem formulation. In Section~\ref{sol_app}, we transform the GC with dynamic codeword assignment problem to a dynamic clustering problem and illustrate its advantage over the original GC and GC with static clustering schemes. Section~\ref{greedy_scheme} presents the proposed greedy dynamic clustering strategy and Section~\ref{num_res} demonstrates its effectiveness through numerical simulations over the existing static GC schemes. Finally, we conclude this paper in Section~\ref{conc} with a summary of the main results along with a discussion of some future directions. 

\section{Preliminaries: Gradient Coding (GC) and Clustering}\label{GC_cl}

In many machine learning problems, given a labeled dataset $\mathcal{D} = \{ (\mathbf{x}_1, y_1), \ldots (\mathbf{x}_s, y_s)\}$, where $\mathbf{x}_1, \ldots, \mathbf{x}_s \in \mathbb{R}^d$ are the data points with corresponding labels $y_1, \ldots, y_s \in \mathbb{R}$, the goal is to solve the following optimization problem
\begin{align}
    \boldsymbol{\theta}^* = \argmin_{\boldsymbol{\theta}\in \mathbb{R}^d} \sum_{i=1}^s l (\mathbf{x}_i, y_i, \boldsymbol{\theta}),
\end{align}
where $l$ is the application-specific loss function and $\boldsymbol{\theta} \in \mathbb{R}^d$ is the parameter vector to be optimized. The optimal parameter vector can be obtained iteratively using GD. The \textit{full gradient} computed over the whole dataset at iteration $t$ is given by $\boldsymbol{g}^{(t)}=\sum_{i=1}^s \nabla l (\mathbf{x}_i, y_i, \boldsymbol{\theta}_{t})$. When the size of the dataset, $s$, is large, the computation of the full gradient becomes a performance bottleneck. To speed up GD iterations, gradient computations can be distributed across multiple workers. However, in many implementations, particularly in the context of `serverless' computing, e.g., Microsoft Azure, Amazon Web Services (AWS), the workers' completion time of assigned tasks can be highly heterogeneous and stochastic over time. In those cases, the overall computation speed of each iteration becomes limited by the slowed \textit{straggling} server. Coded computing techniques tackle the bottleneck due to stragglers by introducing redundant computations in a structured manner such that additional computations carried out by faster servers can compensate for the stragglers. 

\subsection{Gradient Coding (GC)}

GC is a distributed coded computation technique introduced in \cite{UCCT.1} to perform distributed GD across $K$ workers. The complete dataset $\mathcal{D}$ is divided into $K$ non-overlapping equal-size mini-batches, $\mathcal{D}_{1},\ldots,\mathcal{D}_{K}$, and each worker is assigned multiple mini-batches. We denote the set of indices of mini-batches assigned to the $k$th worker with $\mathcal{I}_{k}$, $k \in [K] \triangleq \{1,\ldots,K\}$. Let $\boldsymbol{g}^{(t)}_{k}$ denote the partial gradient for the parameter vector $\boldsymbol{\theta}_{t}$ evaluated over mini-batch $\mathcal{D}_{k}$ at the $t$th GD iteration, i.e.,
\begin{equation}
\boldsymbol{g}^{(t)}_{k}=\frac{1}{\vert \mathcal{D}_{k} \vert} \sum_{(\mathbf{x},y) \in \mathcal{D}_{k}} \nabla l (\mathbf{x}, y, \boldsymbol{\theta}_{t}).
\end{equation}
We note that the \emph{full gradient} is given by $\boldsymbol{g}^{(t)}=\frac{1}{K}\sum^{K}_{k=1}\boldsymbol{g}^{(t)}_{k}$. To tolerate straggling workers, GC assigns redundant mini-batches, and hence, redundant computations, to the workers. 

If a mini-batch $\mathcal{D}_i$ is assigned to worker $k$, i.e., $i \in \mathcal{I}_{k}$, then the corresponding partial gradient $\boldsymbol{g}^{(t)}_{i}$ is computed by the $k$th worker. \textit{Computation load}, $r$, denotes the number of mini-batches assigned to each worker, i.e., $\vert\mathcal{I}_{k}\vert = r$, $\forall k\in[K]$. At each iteration, each worker first computes the $r$ partial gradients, one for each mini-batch available locally, and sends a linear combination of the results, $\boldsymbol{c}^{(t)}_{k}\defeq\mathcal{L}_{k}(\boldsymbol{g}^{(t)}_{i}:i\in\mathcal{I}_{k})$, called a {\em coded partial gradient}. Thus, in the GC scheme, each worker is responsible for computing a single predefined coded partial gradient. The underlying code structure in GC, which dictates the linear combinations formed by each worker, exploits the available redundancy so that the PS can recover the full gradient from only a subset of the combinations. Accordingly, from now on, we refer to the coded partial gradients formed by the workers simply as \emph{codewords.} As shown in \cite{UCCT.1}, the GC scheme can tolerate up to $r-1$ persistent stragglers\footnote{These are the straggler workers that either cannot complete any computation or whose computations are not used while recovering the full gradient \cite{Ozfatura19}.} at each iteration. Formally, for any set of non-straggling workers $\mathcal{W}\subseteq[K]$ with  $ \lvert \mathcal{W}   \rvert = K-r+1 $, there exists a set of coefficients $\mathcal{A}_{\mathcal{W}}=\left \{a^{(t)}_{k}:k\in\mathcal{W}\right\}$ such that
\begin{align}
\sum_{k\in\mathcal{W}} a^{(t)}_{k}\boldsymbol{c}^{(t)}_{k}=\frac{1}{K}\sum^{K}_{k=1}\boldsymbol{g}^{(t)}_{k}.\label{GC_decod}
\end{align}
Thus, at each iteration $t$, the full gradient $\boldsymbol{g}^{(t)}$ can be recovered from any $K-r+1$ codewords.

Next, we present the idea of \textit{clustering} that was introduced in \cite{Ozfatura19} to reduce the average per-iteration completion time of the GC scheme.
\subsection{Gradient Coding with Static Clustering (GC-SC)} \label{GC_cl_static}

In GC with clustering, we divide the workers into $P$ disjoint clusters, each with the same number of workers. Let $\mathcal{K}_{p} \subset [K]$ denote the set of workers in cluster $p$, $p\in[P]$, where $\mathcal{K}_q \cap \mathcal{K}_p = \emptyset$ for $q \neq p$, and $\bigcup_{p\in[P]} \mathcal{K}_p = [K]$. We denote the cluster size by $\ell \triangleq \frac{K}{P}$, where we assume that $K$ is divisible by $P$ for simplicity. The assignment of the workers to the clusters is dictated by an $\ell \times p$ worker assignment matrix, denoted by $\mathbf{A}_{cluster}$, where each column corresponds to a different cluster and the entries in each column correspond to indices of the workers assigned to that cluster. This worker assignment matrix is fixed throughout the training process, hence the name \emph{static clustering.} From now on, we refer to the GC with static clustering scheme as GC-SC.

In GC-SC, each worker is assigned $r$ mini-batches based on its cluster. This is represented by an $r \times k$ data assignment matrix $\mathbf{A}_{data}$, where each column corresponds to a different worker, and the entries in column $i$, $i \in [K]$, represent the mini-batches (correspondingly the partial gradient computations) assigned to the $i$th worker. Equivalently, data assignment can be represented by a $1\times k$ codeword assignment matrix $\mathbf{A}_{code}$, which represents the codewords assigned to the workers, where the codeword assigned to the $i$th worker in the $p$th cluster is denoted by $c_{p,i}$, for $p \in [P]$, $i \in [\ell]$. Let $\mathcal{I}_{\mathcal{K}_{p}}$ denote the set of mini-batches assigned to the workers in the $p$th cluster, i.e., $\mathcal{I}_{\mathcal{K}_{p}}= \bigcup_{k\in\mathcal{K}_p} \mathcal{I}_k$. In GC-SC, the GC scheme is applied to each cluster separately and the workers in cluster $p$ aim at computing
\begin{equation}
\frac{1}{\vert\mathcal{I}_{\mathcal{K}_{p}}\vert}\sum_{k \in \mathcal{I}_{\mathcal{K}_{p}} }\boldsymbol{g}^{(t)}_{k}.
\end{equation}

To illustrate the advantage of the clustering technique, consider $K=12$, $r=2$, and $P=4$. Here, the workers are divided into $4$ clusters, each consisting of $\ell=3$ workers, and each cluster is responsible for computing $3$ of the total $12$ partial gradients. Since $r=2$, each worker aims at computing the assigned $2$ partial gradients. 

In our example, the worker assignment can be specified by the following matrix: 
\begin{align}
\mathbf{A}_{cluster}=
  \begin{bmatrix}
{1} & {2} & {3} & {4}\\
{6} & {7} & {8} & {5}\\
{9} & {10} & {11} & {12}\\
  \end{bmatrix}.\label{A_cluster_st}
\end{align}
In this assignment, workers $1, 6$ and $9$ are in the first cluster, workers $2, 7$ and $10$ are in the second cluster, and so on. The corresponding $\mathbf{A}_{data}$ is given in (\ref{assign_mat_cl}) for the cluster assignment in (\ref{A_cluster_st}). In (\ref{assign_mat_cl}), workers in each cluster are represented by a different color. We use \textcolor{blue}{blue,} \textcolor{red}{red,} \textcolor{magenta}{magenta,} and \textcolor{green1}{green} for clusters $1$, $2$, $3$, and $4$, respectively. The corresponding $\mathbf{A}_{code}$ for the cluster assignment in (\ref{A_cluster_st}) is given in (\ref{codeword_mat}), where, codewords corresponding to different clusters are shown in different colors. Each codeword in $\mathbf{A}_{code}$ is a linear combination of $r=2$ partial gradients. For example, $c_{1,1}$ is a linear combination of partial gradients $g_1$ and $g_2$; $c_{1,2}$ is a linear combination of partial gradients $g_2$ and $g_3$, and $c_{1,3}$ is a linear combination of partial gradients $g_3$ and $g_1$. Thus, given $\mathbf{A}_{cluster}$, either $\mathbf{A}_{data}$ or $\mathbf{A}_{code}$ is sufficient the completely characterize the partial computations that will be carried out by each worker.

\begin{figure}[t]
\centering  \includegraphics[width=0.5\columnwidth]{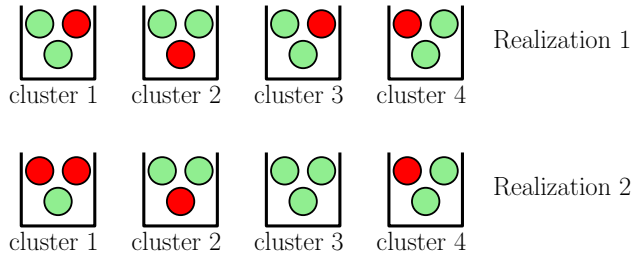}
\caption{Two possible straggler realizations where red and green circles represent the straggling and non-straggling workers, respectively.}
\label{fig_cluster}
\end{figure}

In the original GC scheme, the PS waits until it receives $K-r+1=11$ results at each iteration; hence only $r-1=1$ straggler can be tolerated. With clustering, the PS needs to receive at least $\ell-r+1=2$ results from each cluster to recover the full gradient. Thus, the non-straggling threshold is still $K-r+1$, since more than one straggler cannot be tolerated if they are in the same cluster. However, the non-straggling threshold represents a worst case scenario. With clustering, up to $4$ stragglers can be tolerated if they are uniformly distributed across clusters, i.e., one straggler per cluster, as shown in ``Realization $1$" in Fig.~\ref{fig_cluster}. This shows that, in the case of clustering, the full gradient can be recovered in a much larger set of realizations compared to the original GC scheme. Thus, even if the non-straggling threshold (which corresponds to the worst case scenario) remains the same, clustering will reduce the average per-iteration completion time.

\setcounter{MaxMatrixCols}{20}
\begin{figure*}
\begin{align}
\mathbf{A}_{data}=
  \begin{bmatrix}
  {\color{blue}g_{1}} & {\color{red}g_{4}} & {\color{magenta}g_{7}} &  {\color{green1}g_{10}} &  {\color{green1}g_{11}} & {\color{blue}g_{2}} & {\color{red}g_{5}} & {\color{magenta}g_{8}}  & {\color{blue}g_{3}} & {\color{red}g_{6}} & {\color{magenta}g_{9}}  & {\color{green1}g_{12}}\\
  {\color{blue}g_{2}} & {\color{red}g_{5}} & {\color{magenta}g_{8}}   & {\color{green1}g_{11}} &  {\color{green1}g_{12}} & {\color{blue}g_{3}} & {\color{red}g_{6}} & {\color{magenta}g_{9}}  & {\color{blue}g_{1}} & {\color{red}g_{4}} & {\color{magenta}g_{7}}  & {\color{green1}g_{10}} \\
  \end{bmatrix}\label{assign_mat_cl} 
  \end{align}
\vspace{-8mm}
\end{figure*}
 
\setcounter{MaxMatrixCols}{20}
\begin{figure*}
\begin{align}
\mathbf{A}_{code}=
  \begin{bmatrix}
{\color{blue}c_{1,1}}&{\color{red}c_{2,1}}&{\color{magenta}c_{3,1}}&{\color{green1}c_{4,1}}&{\color{green1}c_{4,2}}&{\color{blue}c_{1,2}}&{\color{red}c_{2,2}}&{\color{magenta}c_{3,2}}&{\color{blue}c_{1,3}}&{\color{red}c_{2,3}}&{\color{magenta}c_{3,3}}&{\color{green1}c_{4,3}}
  \end{bmatrix}\label{codeword_mat}
\end{align}
\vspace{-8mm}
\end{figure*}
 
Formally, with clustering, it is possible to tolerate $r-1$ stragglers in each cluster in the best case scenario, which is when the stragglers are uniformly distributed among the clusters. In this case, it is possible to tolerate $P(r-1)$ stragglers in total. However, this advantage of clustering diminishes in the case of non-uniform distributed stragglers among the clusters, which may be the case in practice. As shown in ``Realization $2$" in Fig.~\ref{fig_cluster}, even though there are still $8$ non-straggling workers, the PS cannot compute the full gradient (in the case of persistent stragglers) when the stragglers are not uniformly distributed across the clusters. To this end, in the next section, we introduce the concept of dynamic codeword assignment, which dynamically changes codewords computed by the workers at each iteration based on the past straggler behavior to further improve the performance of the clustering technique.

\section{GC with Dynamic Codeword Assignment}\label{GC_dyn}

In the conventional coded computation approaches, including the GC, the assignment of the dataset to the workers and the code to be used are static and set at the beginning of the training process. That is, at every iteration, a worker tries to compute the gradient estimates for all the mini-batches assigned to it, and returns their exact same linear combination to the PS. Thus, in order to recover the desired computation result at each iteration, the codes are designed for the worst case scenario. The core idea behind dynamic codeword assignment is to change the codewords assigned to the workers dynamically based on the observed straggling behavior. Dynamic codeword assignment is driven by two policies; namely, \textit{data assignment} and \textit{codeword assignment}. The data assignment policy, denoted by $\Pi_{d}$, is executed only once at the beginning of training and assigns up to $m$ mini-batches to each worker, where $m$ denotes the memory constraint, i.e., 
\begin{equation}
\Pi_{d}:  \mathcal{D} \mapsto \left\{\mathcal{I}_{1},\ldots,\mathcal{I}_{K}: \vert\mathcal{I}_{k} \vert \leq m\right\}.
\end{equation}
We note that even though each worker can be allocated up to $m$ mini-batches, each will compute only $r$ of them at each iteration; hence, the computation load at each iteration remains the same. On the other hand, we can have $m \choose r$ codewords that can be assigned to each worker depending on which subset of $r$ computations it carries out among $m$ possibilities. Here, we introduce $\mathcal{C}=\left\{\mathcal{C}_{1},\ldots,\mathcal{C}_{K}\right\}$, where $\mathcal{C}_{k}$ denotes the set of feasible codewords corresponding to dataset $\mathcal{I}_{k}$. That is, $\mathcal{C}_{k}$ denotes the set of codewords that may be assigned to the $k$th worker at each iteration, where each codeword is a linear combination of $r$ gradient estimates that can be computed by this worker. 

We would like to highlight that with dynamic codeword assignment, the PS will specify at each iteration which codeword must be computed by each worker. This introduces additional communication requirement compared to the static schemes, such as GC and GC-SC. On the other hand, this information can be piggybacked on other control information that must be communicated from the PS to the workers at each iteration, such as signalling the end of an iteration and the transmission of the updated model parameters. However, it is still important to keep this additional information minimal by designing a codebook with minimal $|\mathcal{C}_k|$.

At the beginning of each iteration $t$, codeword assignment policy $\Pi_{a}$ is executed by the PS based on the past straggler behavior of the workers up to iteration $t$, $\mathbf{S}^{[t-1]}$, i.e.,
\begin{equation}
\Pi^{(t)}_{a}(\mathbf{S}^{[t-1]}, \Pi_d):  \mathcal{C}\mapsto {\mathbf{c}}^{t}= \left\{c^{t}_{1},\ldots,c^{t}_{K}\right\},\label{assignment}
\end{equation}
where $c^{t}_{k} \in \mathcal{C}_k$ is the codeword assigned to the $k$th worker at iteration $t$ and $\mathbf{S}^{[t-1]} \triangleq (\mathbf{S}^1, \ldots, \mathbf{S}^{t-1})$, while $\mathbf{S}^t=(S^t_1,\ldots, S^t_K)$ denotes the straggler behavior at each iteration $t$, where $S^t_{k}=0$ if the $k$th worker is a straggler at iteration $t$, and $S^t_{k}=1$ otherwise.\footnote{In this work, we assume an on/off straggling behavior for each worker such that a worker's straggling status can change over iterations. Workers can still deliver computation results in the straggling state but their computations are much slower. This type of two-state straggling behavior is observed in empirical studies over Amazon EC2 clusters \cite{UCCT.1, Yang19}.} 

The completion time of iteration $t$ for a given data assignment policy $\Pi_d$ depends on the codeword assignment ${\mathbf{c}}_{t}$ and the straggler realization $\mathbf{S}^{t}$. Here, our objective is to minimize the expected completion time of each iteration based on the past straggler behavior for a given $\Pi_d$:
\begin{equation}
\min_{\Pi^{(t)}_{a}} \mathbb{E}_{\mathbf{S}^{t}\vert \mathbf{S}^{[t-1]}, \Pi_d} Q({\mathbf{c}}^{t},\mathbf{S}^{t}),
\end{equation}
where $Q_{}({\mathbf{c}}^{t},\mathbf{S}^{t})$ is the completion time of iteration $t$ under codeword assignment ${\mathbf{c}}_{t}$ and the straggler realization $\mathbf{S}^{t}$.\\
\indent We remark that the codeword assignment policy $\Pi^{(t)}_{a}$ highly depends on the data assignment policy $\Pi_{d}$ since in most of the coded computation scenarios the data assignment policy is driven by the employed coding strategy. Thus, designing a data assignment policy $\Pi_{d}$ without any prior knowledge on the coding strategy is a challenging task. To this end, in the next section, we reformulate the dynamic codeword assignment problem where the coding strategy, consequently the set of codewords, are fixed at the beginning and data assignment is performed based on the underlying coding strategy.

\section{GC with Dynamic Clustering (GC-DC)}\label{sol_app}

In this section, we reformulate the dynamic codeword assignment problem, and introduce the GC-DC scheme. For the construction of the GC-DC scheme, we perform three steps; namely, codeword construction, codeword distribution, and dynamic clustering, where the first two steps are executed once at the beginning of training and the last one is executed at each iteration. Our code construction will be based on GC-SC presented in Section~\ref{GC_cl_static}, and we will transform the dynamic codeword assignment problem into a dynamic clustering problem. We note that the number of clusters $P$ is fixed and decided at the beginning of the training.

\subsection{Codeword Construction}

In the GC-DC scheme, we will request each worker to compute and return a codeword at each iteration. Remember that each codeword is a specified linear combination of the gradient estimates for a subset of $r$ mini-batches, and the PS and the workers need to agree on how to form these linear combinations in advance. Here, the set of codewords $\mathcal{C}$ is a union of smaller disjoint codeword sets, i.e., $\mathcal{C}=\bigcup^{P}_{p=1}\mathcal{C}^{p}$, such that the codewords in each set $\mathcal{C}^{p}$, $p\in [P]$, are encoded and decoded independently and correspond to a particular cluster. For example, in (\ref{codeword_mat}), $\mathcal{C}^{1}=\{c_{1,1}, c_{1,2}, c_{1,3}\}$, where $\mathcal{C}^{1}$ is disjoint from the rest of the codeword set.

\subsection{Codeword Distribution}\label{sect:code_dist}

The codewords in $\mathcal{C}$ are distributed among the workers according to a policy $\Pi_{c}$, i.e.,
\begin{equation}
\Pi_{c}(\mathcal{C}): \mathcal{C} \mapsto \left\{\mathcal{C}_{1},\ldots,\mathcal{C}_{K}\right\},\label{codeword_pol} 
\end{equation}
where we remark that $\mathcal{C}_{k}$ denotes the set of codewords that can be assigned to the $k$th worker at each iteration. Now, let $\mathcal{I}(c)\subseteq\mathcal{D}$ be the minimal subset of mini-batches that is sufficient to construct codeword $c$, where we have $|\mathcal{I}(c)|\leq r$. Given the codeword distribution policy $\Pi_{c}$, any feasible data assignment policy $\Pi_{d}$ should satisfy the following constraint
\begin{equation}
\mathcal{I}_{k}\supseteq\bigcup_{c\in\mathcal{C}_{k}}\mathcal{I}(c), \quad \forall k \in [K].\label{assignment_pol}
\end{equation}
Based on this constraint, we observe that, given $\Pi_{c}(\mathcal{C})$, the minimum memory is used when $\mathcal{I}_{k}=\bigcup_{c\in\mathcal{C}_{k}}\mathcal{I}(c), \forall k\in [K]$. Thus, we note that the data assignment policy $\Pi_d$ is determined according to the codeword distribution policy $\Pi_c$. In other words, we first perform codeword distribution and then assign the corresponding mini-batches to the workers.

Next, we describe the codeword distribution policy $\Pi_{c}$ in (\ref{codeword_pol}) for the proposed GC-DC scheme. We first assign each worker to $n$ clusters. Each cluster $p$ corresponds to a set of codewords $\mathcal{C}^p$ with $|\mathcal{C}^p|=\ell$. We say that a worker is in cluster $p$, if that worker is assigned all $\ell$ codewords in $\mathcal{C}^p$. Hence, in the proposed scheme, each worker is assigned codewords from an $n$-subset of $\{\mathcal{C}^1,\ldots,\mathcal{C}^P\}$.\footnote{That is, under the proposed GC-DC scheme, we have $|\mathcal{C}_k|=n\ell$ such that each worker may be assigned all $\ell$ codewords for each of the clusters that it belongs to.} With this, we form a worker cluster assignment matrix $\mathbf{A}_{cluster}$ of size $\ell n \times P$. Here, the $p$th column of $\mathbf{A}_{cluster}$ illustrates the workers assigned to the $p$th cluster, where $w_k$ denotes the $k$th worker, $k\in [K]$. One such example $\mathbf{A}_{cluster}$ for our continuing example is given in (\ref{A_cluster}) for $n=2$. 
\begin{align}
\mathbf{A}_{cluster}=
  \begin{bmatrix}
w_{1} & w_{2} & w_{3} & w_{4}\\
w_{6} & w_{7} & w_{8} & w_{5}\\
w_{9} & w_{10} & w_{11} & w_{12}\\
w_{4} & w_{1} & w_{2} & w_{3}\\
w_{7} & w_{8} & w_{5} & w_{6}\\
w_{10} & w_{11} & w_{12} & w_{9}\\
  \end{bmatrix}.\label{A_cluster}
\end{align}

When assigning workers to clusters, we start by dividing workers into $P$ groups according to their indices. For example, in our continuing example for $P=4$ and $K=12$, these groups are $\{w_1, \ldots, w_4\}$, $\{w_5, \ldots, w_8\}$, and $\{w_9, \ldots, w_{12}\}$. Then, we utilize a circular shift operator and sample $n$ shift amounts in $\{0,\ldots,P-1\}$ uniformly at random without replacement for each of these groups. We circularly shift each of these groups according to the corresponding sampled shift amounts and form the worker cluster assignment matrix $\mathbf{A}_{cluster}$. For example, in the first and fourth rows of (\ref{A_cluster}), the shift amounts for workers $\{w_1,\ldots, w_4\}$ are $0$ and $1$, respectively. As a result of these random shifts, worker $w_1$ is assigned to the first and second clusters, worker $w_2$ is assigned to the second and third clusters, and so on. Similarly, from the second and fifth rows of (\ref{A_cluster}), we observe that the shift amounts for workers $\{w_5,\ldots, w_8\}$ are $3$ and $2$, respectively. We note that, since the random shifts for the same set of workers, e.g., workers $\{w_1,\ldots, w_4\}$, are sampled without replacement, each worker is assigned to exactly $n=2$ distinct clusters.

We remark that, given $n$, the memory requirement $m$ of the proposed GC-DC scheme is given by $m=n\ell$. Thus, for $n=2$ and $\ell=3$, each worker stores $6$ mini-batches in this example. 

By constructing $\mathbf{A}_{cluster}$, we essentially perform the codeword distribution as each worker is assigned all $\ell$ codewords for each of the $n$ clusters that it is associated with. For example, from (\ref{A_cluster}) we deduce that worker $1$ has all the codewords in sets $\mathcal{C}^1$ and $\mathcal{C}^2$, i.e., $\mathcal{C}_1 = \mathcal{C}^1 \cup \mathcal{C}^2 = \{c_{1,1}, c_{1,2}, c_{1,3}, c_{2,1}, c_{2,2}, c_{2,3}\}$. With this, we perform the data assignment and assign corresponding mini-batches to each worker to form the data assignment matrix such that the constraint in (\ref{assignment_pol}) is satisfied with equality. Correspondingly, $\mathcal{I}_1 = \{\mathcal{D}_1,\ldots,\mathcal{D}_6\}$ so that worker $1$ can compute partial gradients $g_1,\ldots,g_6$ to form any one of these $6$ codewords. 

\subsection{Dynamic Clustering}

The key idea behind dynamic clustering is to associate each worker to more than one cluster by assigning more than $r$ mini-batches to each worker. Assuming that a worker is associated with $n$ clusters, each worker is assigned a total of $n\ell$ codewords so that a worker can replace any worker in the $n$ clusters it is associated with by computing a codeword that would be computed by the worker to be replaced in the original GC scheme with clustering. Then, at each iteration the PS selects one of the $n\ell$ codewords for each worker based on the previous straggler realization through a codeword assignment policy $\Pi_a$ given in (\ref{assignment}). We note that, even though more than one codeword is assigned to each worker, computation load is still $r$ as in the original GC scheme, and each worker still computes only one codeword consisting of $r$ partial gradient computations at each iteration. 


To see the benefit of the proposed GC-DC scheme, we consider $\mathbf{A}_{cluster}$ and corresponding codewords for a particular straggler realization $\mathbf{S}=[{\color{blue}1},{\color{red}1},{\color{magenta}0},{\color{green1}1},{\color{green1}1},{\color{blue}1},{\color{red}1},{\color{magenta}0},{\color{blue}0},{\color{red}1},{\color{magenta}0},{\color{green1}1}]$, where, colors follow the cluster assignment in the static clustering case, i.e., $\mathbf{A}_{cluster}$ given in (\ref{A_cluster_st}). Under the GC-SC scheme, it is not possible to recover partial gradients corresponding to the third cluster as we do not have $\ell-r+1=2$ non-straggling workers in that cluster.\footnote{We note that this is the case assuming straggling workers do not return any computation results. Even if they do, whenever there are less than $\ell-r+1$ non-straggling workers in a cluster, the PS has to wait for at least one of the straggling workers to return its computation which may incur a significant delay in the completion time of that iteration.} Moreover, if this straggling behavior persists for a substantial duration of time, the overall computation time will suffer drastically. To mitigate this, in the case of dynamic clustering, we observe in (\ref{A_cluster}) that worker $w_5$ can replace worker $w_3$ since it can compute codeword $c_{3,1}$ which is the codeword that was originally assigned to worker $w_3$ in (\ref{codeword_mat}) in the GC-SC scheme. This does not affect the recoverability of the partial gradients assigned to the fourth cluster, to which worker $w_5$ initially belongs, since that cluster has $2$ more non-straggling workers, workers $w_4$ and $w_{12}$. Further, worker $w_2$ can replace worker $w_8$ so that all partial gradients can be recovered successfully. Equivalently, we have assigned the clusters such that non-straggling workers $w_2$ and $w_5$ now belong to the $3$rd cluster by ensuring that all other clusters still have at least $\ell-r+1=2$ non-straggling workers. Thus, dynamic clustering increases the set of straggler realizations for which the full gradient recovery is possible compared to static clustering. 

Since each worker can replace any worker in all the $n$ clusters that it is assigned to, we essentially form the clusters, dynamically at each iteration through codeword assignments, hence the name \emph{dynamic clustering.} That is, based on the codeword distribution presented in Section~\ref{sect:code_dist}, we can assign $\ell$ workers to each cluster according to the given worker cluster assignment matrix $\mathbf{A}_{cluster}$ without explicitly stating which worker will compute which codeword. With this, our aim is to dynamically form clusters at each iteration to minimize the average completion time of an iteration given the past straggler behavior and the worker-cluster assignment matrix $\mathbf{A}_{cluster}$. 


Next, we characterize the average completion time of an iteration for a given cluster assignment. We denote the $k$th smallest of random variables $Y_1, \ldots, Y_n$ as $Y_{k:n}$. The completion time of iteration $t$ for cluster $p$ is given by the time the PS receives the earliest $\ell-r+1$ results from that cluster such that
\begin{align}
    Q^p(\mathbf{c}^t, \mathbf{S}^{t}) = \{X^p_{1,r}, \ldots, X^p_{\ell,r}  \}_{\ell-r+1:\ell}, \quad p \in [P], \label{cluster_comp}
\end{align}
where $\mathbf{c}^t$ is the set of codewords assigned to the workers as in (\ref{assignment}) and $X^p_{k,r}$, $k \in [\ell]$, is the computation duration of the $k$th worker of cluster $p$, i.e., the time it takes for that worker to compute $r$ partial gradients. Noting that iteration $t$ ends when each cluster recovers its corresponding partial gradients, completion time of iteration $t$ is given by
\begin{align}
    Q(\mathbf{c}^t, \mathbf{S}^{t}) = \max_{p\in[P]} Q^p(\mathbf{c}^t, \mathbf{S}^{t}). \label{iter_comp}
\end{align}

Since some of the workers are stragglers, computation capabilities of the workers are not identical. In this case, minimizing the iteration completion time given in (\ref{iter_comp}) through cluster assignments is not an analytically tractable problem. Instead, in the next section, we propose a greedy dynamic clustering strategy that aims to uniformly place stragglers across clusters at each iteration to speed up GC.

\section{Greedy Dynamic Clustering Strategy }\label{greedy_scheme}

In line with the observations on Amazon EC2 instances in \cite{UCCT.1, Yang19}, in this section, we consider a stochastic straggling behavior for the workers. In particular, we assume that workers' computation statistics are independent from each other, and follow a two-state Markov process. That is, at each iteration a worker can be either in a straggling or a non-straggling state. Once a worker starts straggling, it operates significantly slower than the non-straggling performance and remains straggling for a while. This may model an increased load at a worker for a period of time, which reduces the computational resources that can be allocated for the specific computation task. Our proposed greedy algorithm utilizes this time-correlated straggling behavior to assign straggling workers to different clusters. At each iteration, the PS identifies the stragglers based on the past observations and implements a greedy dynamic clustering strategy to uniformly distribute the stragglers across clusters to improve the completion time of each iteration. We note that the performance gain of the proposed GC-DC scheme is prominent when the computation speeds of the workers are not identically distributed over iterations, e.g., they exhibit time-correlated straggling behavior, as the GC-DC scheme gains from adapting to the straggling behavior by carefully placing the workers to clusters at each iteration. 

Inspired by the bin packing problem \cite{Korte08}, we consider clusters as bins and workers as balls as in Fig.~\ref{fig_cluster}. Unlike the bin packing problem, which aims to place balls of different volumes into a minimum number of bins of finite volume, in our setting, the number of bins (clusters) is fixed and our aim is to distribute the straggling workers as uniformly as possible to clusters using the worker cluster assignment matrix $\mathbf{A}_{cluster}$. Our dynamic clustering algorithm has two phases: in the first phase, based on the previous straggler realization, we place straggler and non-straggler workers into clusters separately following a specific order, and in the second phase, any placement conflict that may happen in the first phase (i.e., if a worker cannot be placed into any of the remaining clusters) is resolved through worker swap between the corresponding clusters. During worker placement, clusters take turns based on a specified order and we implement a greedy policy such that, once its turn comes, each cluster selects the first available worker that can be assigned to that cluster based on the given worker cluster assignment matrix $\mathbf{A}_{cluster}$.

In what follows we describe in detail the proposed dynamic clustering strategy, which is also presented in Algorithm~\ref{alg:greedy}. Given the worker cluster assignment matrix $\mathbf{A}_{cluster}$, without loss of generality, we first reorder workers in each cluster according to their indices such that 
\begin{align}
    \mathbf{A}_{cluster}(i,p) < \mathbf{A}_{cluster}(j,p), \quad i < j, \quad p \in [P],\label{order}
\end{align}
where $\mathbf{A}_{cluster}(i,p)$ denotes the index of the worker in the $i$th position in cluster $p$. For example, in $\mathbf{A}_{cluster}$ given in (\ref{A_cluster}), $\mathbf{A}_{cluster}(1,2)$ is $2$ since it corresponds to worker $w_2$. Once its turn comes, each cluster starts selecting workers with the lowest indices first. We note that, if the workers have heterogeneous computing capabilities, then in this step we order workers according to their speed of computation, such that the fastest workers are selected first, which we will consider in Section~\ref{num_res_het}. For ease of exposition, here, we provide the algorithm when all the straggling workers have identical computation statistics, and similarly all the non-straggling workers have the same computation statistics with each other. Therefore, there is no preference among workers within each group, and ordering them according to their indices is appropriate.

We assume that at the end of each iteration, each worker accurately detects its straggling status and informs the PS using an instantaneous feedback. The straggling state information is in general not available to the worker before that iteration ends due to the unpredictable and highly varying nature of computing resources in distributed computing systems. Since the current straggling behavior is random following the underlying Markov process, at iteration $t$, the algorithm starts by deducing the sets of non-straggling and straggling workers $\mathcal{K}_f$ and $\mathcal{K}_s$ from $\mathbf{S}^{t-1}$. We note that, at each iteration, $\mathcal{K}_f \cup \mathcal{K}_s = [K]$. The proposed algorithm uses the straggler statistics from iteration $t-1$ to perform dynamic clustering at iteration $t$, which makes this algorithm suitable for Markovian straggling models.

\subsection{Phase I - Worker Placement}

We place straggling and non-straggling workers separately to the clusters following a specific order. If the number of non-straggling workers is higher than the stragglers, i.e., $|\mathcal{K}_f| \geq |\mathcal{K}_s|$, we start by placing the non-stragglers and vice-versa. 

For the sake of demonstration, we assume $|\mathcal{K}_f| \geq |\mathcal{K}_s|$ and place the non-straggling workers first. Let $O_f$ denote the order in which the clusters select workers such that $O_f(p)$ gives the order in which the $p$th cluster selects workers. To determine the exact order, we define $\mathcal{K}^p_f$ and $\mathcal{K}^p_s$, which denote the set of non-straggling and straggling nodes that can be assigned to cluster $p$, respectively. We remark that worker $k$ can be assigned to cluster $p$ if it is in column $p$ of $\mathbf{A}_{cluster}$, i.e., $w_k \in \mathbf{A}_{cluster}(:,p)$. With this, we determine the order vector such that
\begin{align}
    O_f(p) < O_f(\bar{p}) \text{ if } |\mathcal{K}^p_f| < |\mathcal{K}^{\bar{p}}_f| \quad  p, \bar{p} \in [P].
\end{align}
That is, clusters with less availability select workers first. In the case of equal availability, i.e., $|\mathcal{K}^p_f| = |\mathcal{K}^{\bar{p}}_f|$, cluster with the smaller index selects first, i.e., $O_f(p) < O_f(\bar{p})$ for $p<\bar{p}$. The order for straggler placement $O_s$ is determined accordingly using $\mathcal{K}^{p}_s$, for $p\in[P]$.

Once the order $O_f$ is determined, non-straggling workers are placed into clusters following $O_f$. As stated in lines 16-23 of Algorithm~\ref{alg:greedy}, once its turn comes, each cluster $p$ with an open spot, i.e., each cluster $p$ that currently has less than $\ell$ workers, selects the first available non-straggling worker from $\mathbf{A}_{cluster}(:,p)$, $p\in [P]$. Once a non-straggling worker is assigned to a cluster, we remove it from $\mathcal{K}_f$ and $\mathbf{A}_{cluster}$. We note that this assignment continues until there is no unassigned non-straggling worker is left in $\mathcal{K}_f$ or a placement conflict is observed. Then, the straggler workers are placed following a similar procedure with the order vector $O_s$.

During Phase I, the algorithm makes at most $M$ such placement attempts, where $M>0$ is a sufficiently large number. If after $M$ turns, a worker cannot be assigned to any of the remaining clusters, we deduce that there is a placement conflict and move on the second phase of the algorithm.

\subsection{Phase II - Conflict Resolution}

Assume that there is a placement conflict at the end of Phase I such that worker $k$ cannot be placed to the remaining cluster $p$. That is, all of the $n$ clusters that worker $k$ can be assigned to are full, i.e., already have $\ell$ workers, and cluster $p$ needs one more worker. In such a case, the second conflict resolution phase of the algorithm starts. 

Let $\mathcal{P}_k$ denote the set of possible clusters for worker $k$ such that $|\mathcal{P}_k|=n$. In the conflict resolution step, as stated in lines 26-35 of Algorithm~\ref{alg:greedy}, we look for a worker $\bar{k}$, which has been assigned to one of the clusters in $\mathcal{P}_k$ in Phase I such that $w_{\bar{k}} \in \mathbf{A}_{cluster}(:,p)$. That is, even though worker $\bar{k}$ has been assigned to cluster $\bar{p} \in \mathcal{P}_k$ during Phase I, it can be assigned to cluster $p$ as well. Once we detect first such worker, we swap its position with worker $k$. That is, we assign worker $k$, the conflicted worker, to cluster $\bar{p}$ and worker $\bar{k}$ to cluster $p$, the conflicted cluster.

We note that there might be multiple placement conflicts at the end of Phase I, in which case the conflict resolution step is repeated until all cases are resolved.

\begin{algorithm}[t]
\caption{Proposed dynamic clustering strategy}\label{alg:greedy}
\begin{algorithmic}[1]
    \State Given $\mathbf{A}_{cluster}$, $K$, $P$, $n$, $\mathbf{S}^{0}$ such that w.l.o.g. $\mathbf{A}_{cluster}(i,p) < \mathbf{A}_{cluster}(j,p) $ for $i < j$, $p \in [P]$
   \For{$t=1,\ldots,T$}
        \State Observe $\mathbf{S}^{t-1}$ and deduce $\mathcal{K}_f$ and $\mathcal{K}_s$, i.e., sets of non-straggling and straggling workers in iteration $t-1$
        \State\underline{\textbf{Phase I:}}
        \State Place workers to clusters following an order
        \If{$|\mathcal{K}_f| \geq |\mathcal{K}_s|$}
            \State Place non-stragglers first
        \Else
            \State{Place stragglers first}
        \EndIf
       \State\underline{\textbf{Phase II:}}
       \State Conflict resolution in the case of an assignment problem in Phase I
    \EndFor
\State\underline{\textbf{Order determination:}}
\State $O_f(p) < O_f(\bar{p})$ if $|\mathcal{K}^p_f| < |\mathcal{K}^{\bar{p}}_f|$ or ($|\mathcal{K}^p_f| = |\mathcal{K}^{\bar{p}}_f|$ and $p<\bar{p}$) for $p, \bar{p} \in [P]$ 
\State Use $O_s$ in the case of straggler placement with $\mathcal{K}^{p}_s$ for $p \in [P]$
\State\underline{\textbf{Non-straggler placement:}}
\State $i = 1$
\While {$|\mathcal{K}_f| >0$ and $i<M$}
\State $j = \mod(i, P)$ with $j \gets P$ when $\mod (i, P) = 0$
\State Cluster to assign is $\bar{p}$ such that $O_f(\bar{p})=j$
\If{ $size(\text{cluster } \bar{p}) < \ell$}
\State Assign the first non-straggling worker from $\mathbf{A}_{cluster}(:,\bar{p})$ to cluster $\bar{p}$
\State Remove the assigned worker from $\mathcal{K}_f$ and $\mathbf{A}_{cluster}$
\EndIf
\State $i = i+1$
\EndWhile
\State\underline{\textbf{Straggler placement:}}
\State Follow steps 16-23 using $\mathcal{K}_s$ and $O_s$
\State\underline{\textbf{Conflict resolution:}}
\State Given a conflicted worker $k$ and corresponding conflicted cluster $p$
\State Identify the clusters $\mathcal{P}_k$ that worker $k$ can be assigned to such that $|\mathcal{P}_k|=n$
\State{$i=1$}
\While{Worker $k$ is not assigned to any cluster}
\State Select cluster $\bar{p}$ such that $\bar{p}=\mathcal{P}_k(i)$
\If{There is a worker $\bar{k}$ in cluster $\bar{p}$ such that $w_{\bar{k}} \in \mathbf{A}_{cluster}(:,p)$}
\State Assign worker $\bar{k}$ to cluster $p$
\State Assign worker $k$ to cluster $\bar{p}$
\EndIf
\State{$i=i+1$}
\EndWhile
\end{algorithmic}
\end{algorithm}

To illustrate the proposed worker replacement policy in detail, we consider the cluster assignment matrix in (\ref{A_cluster}), and without loss of generality, order workers in an increasing order in each column to obtain
\begin{align}
\mathbf{A}_{cluster}=
  \begin{bmatrix}
w_{1} & w_{1} & w_{2} & \color{red}{w_{3}}\\
w_{4} & w_{2} & \color{red}{w_{3}} & w_{4}\\
\color{red}{w_{6}} & \color{red}{w_{7}} & \color{red}{w_{5}} & \color{red}{w_{5}}\\
\color{red}{w_{7}} & \color{red}{w_{8}} & \color{red}{w_{8}} & \color{red}{w_{6}}\\
w_{9} & w_{10} & w_{11} & w_{9}\\
w_{10} & w_{11} & w_{12} & w_{12} \label{ex_cluster}
  \end{bmatrix},
\end{align}
where the straggling workers are shown in red. The straggler realization for this example is $\mathbf{S}=[1,1,0,1,0,0,0,0,1,1,1,1]$. Here, we have $5$ straggling and $7$ non-straggling workers, i.e., $|\mathcal{K}_s|=5$ and $|\mathcal{K}_f|=7$.

\begin{figure}[t]
	\centering  \includegraphics[width=0.6\columnwidth]{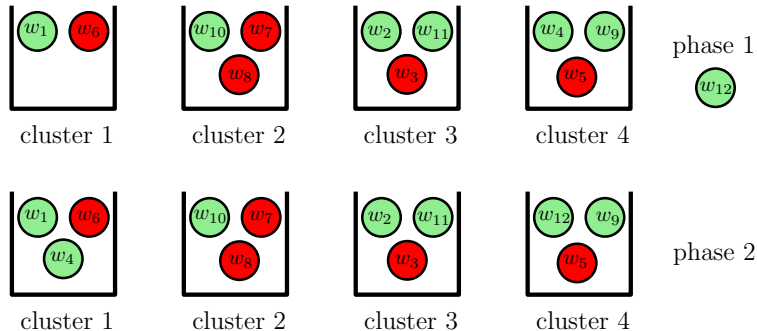}
	\caption{The proposed worker placement strategy.}
	\label{fig_propstrat}
\end{figure}

Since there are more non-straggling workers than stragglers, we place the non-straggling workers first. To determine a non-straggling worker placement order, we find the number of available non-straggling workers in each cluster. One can observe in (\ref{ex_cluster}) that, cluster $1$ and cluster $2$ have $4$ available non-straggling workers that can be assigned to these clusters whereas cluster $3$ and cluster $4$ have $3$ available non-straggling workers. That is, we have $|\mathcal{K}^1_f|=|\mathcal{K}^2_f|=4$ and $|\mathcal{K}^3_f|=|\mathcal{K}^4_f|=3$. Based on these, we deduce a placement order $O_f = [3,4,1,2]$ such that clusters take turns based on this placement order.\footnote{In a more refined implementation, this order can dynamically change after each round of worker placement, i.e., after all clusters select one worker, to better reflect the clusters with less availability as worker placement continues.} At each turn of a particular cluster, a single worker is assigned to that cluster according to the aforementioned greedy policy. In our example, we start with the third cluster and $w_2$ is assigned to this cluster. Then, the fourth cluster gets $w_4$ and so on. This process continues until all the non-straggling workers are placed into clusters (or until a placement conflict is observed). If a cluster is assigned $\ell=3$ workers, we say that cluster is full and do not assign any more workers to that cluster.
Next, we determine the placement order of straggling workers in a similar fashion. One can deduce from (\ref{ex_cluster}) that the order of placement for the stragglers is $O_s=[1,2,3,4]$ as clusters $1$ and $2$ have the least availability. Based on this order, stragglers are also placed using the greedy policy described above and the first phase terminates with the worker placement shown in Fig.~\ref{fig_propstrat}. Here, we observe a placement conflict as $w_{12}$ has not been assigned to any cluster whereas cluster $1$ needs one more worker, but $w_{12}$ cannot be assigned there.

We start the second phase of the proposed worker placement algorithm to place $w_{12}$ into a cluster that has a worker which can be assigned to the first cluster. We see from (\ref{ex_cluster}) that $w_{12}$ can be assigned to clusters $3$ or $4$. None of the workers which has been assigned to cluster $3$ in Phase I can be assigned to the first cluster. Then, the algorithm looks as cluster $4$ and identifies that $w_4$, which has been assigned to the fourth cluster in the first phase, can go to the first cluster. With this, we swap workers $w_4$ and $w_{12}$, which yields the final placement in Fig.~\ref{fig_propstrat}. 

At the end of the algorithm we see that the stragglers are placed into the clusters as uniformly as possible: cluster $2$ has two stragglers while the remaining clusters have only $1$ straggler each. We note that since we have only $7$ non-straggling workers, less than the worst case scenario of $P(\ell-r+1)=8$ non-stragglers, the full recovery is possible for the static clustering scheme. Thus, the proposed dynamic clustering scheme does not improve the worst case scenario. Rather, it speeds up the GC scheme by uniformly placing the stragglers across clusters. This process is repeated at each iteration to dynamically change the clusters based on the straggler observations. 

We note that at the end of the first phase, there are $4$ other workers, namely workers $w_4, w_7, w_9$, and $w_{10}$, that can be placed into the first cluster, which had placement conflict at the end of Phase I of the algorithm. Even if $\ell=3$ of them would have been assigned to cluster $2$, which worker $w_{12}$ cannot be assigned, the remaining one of them still would have been assigned to either cluster $3$ or $4$. Thus, it is guaranteed that cluster $3$ and cluster $4$ have at least one worker that can be assigned to cluster $1$ so that the placement conflict can be resolved. The next lemma formally states this guarantee.

\begin{lemma}
    Assume that we have a conflicted worker $k$ which cannot be assigned to the remaining cluster $p$ in Phase I. Then, if
    \begin{align}
        n > \frac{P(K-1)}{2K},\label{cond}
    \end{align}
    it is guaranteed that at least one worker in one of the clusters in $\mathcal{P}_k$ can be assigned to cluster $p$ so that the placement conflict can be resolved.
\end{lemma}

\begin{Proof}
In the proof we consider the worst case scenario such 
that $\ell-1$ workers have already been assigned to cluster $p$ in Phase I. Thus, in the remaining $P-1$ clusters other than cluster $p$, there are $n\ell-\ell+1$ workers that can be assigned to cluster $p$. We want to make sure that, at the end of Phase I of the algorithm, at least one of those workers is assigned to a cluster in set $\mathcal{P}_k$, which, as previously stated, denotes the set of clusters that worker $k$, the conflicted worker, can be assigned to. Except cluster $p$, there are $P-n-1$ clusters that worker $k$ cannot be assigned to. These $P-n-1$ clusters can at most have $(P-n-1)\ell$ workers after Phase I. Thus, as long as 
\begin{align}
    n\ell-\ell+1 > (P-n-1)\ell,
\end{align}
there is at least one worker that can be assigned to cluster $p$ in one of the clusters in $\mathcal{P}_k$, which yields (\ref{cond}) since $\ell = \frac{K}{P}$.
\end{Proof}
In the previous example, (\ref{cond}) is satisfied since $K=12$, $P=4$, and $n=2$ such that $n>\frac{11}{6}$.

In the next section, we analyze the performance of this dynamic clustering strategy through numerical simulations.

\section{Numerical Results}\label{num_res}

In this section, we provide numerical results comparing the proposed GC-DC scheme with GC-SC as well as the original GC scheme using a model-based scenario for computation latencies. For the simulations, we consider a linear regression problem over synthetically created training and test datasets, as in \cite{CC.4}, of sizes $2000$ and $400$, respectively. We set the size of the model to $d = 1000$. A single simulation consists of $T=400$ iterations. For all the simulations, we use learning rate $\eta = 0.1$. To model the computation delays at the workers, we adopt the commonly used shifted exponential model \cite{entropy}, and assume that the probability of completing $r$ partial gradient computations at worker $k$ by time $t$ is given by
\begin{equation}\label{exp_compute}
  \mathbb{P}[X_{k,r} \leq t] \triangleq
  \begin{cases}
    1-e^{-\mu_k(\frac{t}{r}-\alpha_k)}, & \text{if $t \geq r\alpha_k$}, \\
    0, & \text{otherwise},
  \end{cases}
\end{equation}
where $\alpha_k>0$ is a constant shift indicating that a single computation duration cannot be smaller than $\alpha_k$ and $\mu_k>0$ denotes the straggling effect.
We consider two different models for the time-correlated straggling behavior: the homogeneous and heterogeneous worker models, which we discuss next.

\subsection{Gilbert-Elliot Model with Homogeneous Workers}\label{homogenous_workers}
We model the straggling behavior of the workers based on a two-state Markov chain: a slow state $s$ and a fast state $f$, such that computations are completed faster when a worker is in state $f$. Specifically, in (\ref{exp_compute}) we have rate $\mu_f$ in state $f$ and rate $\mu_s$ in state $s$, where $\mu_f > \mu_s$ as in \cite{Ozfatura20, Buyukates20a}. That is, each worker has two possible rates based on its straggling statistics. We assume that the state transitions only occur at the beginning of each iteration with probability $p$; that is, with probability $1-p$ the state remains the same. A low switching probability $p$ indicates that the straggling behavior tends to remain the same in consecutive iterations with occasional transitions. We set $p=0.05$, $\alpha = 0.01$, $\mu_s = 0.1$, and $\mu_f = 10$. We assume that the transition probability $p$ along with the computation rates $\mu_s$ and $\mu_f$ are known to the PS. At the end of each iteration, workers inform the PS regarding their straggling status before the next iteration starts. With this information along with the knowledge of transition probability $p$, the PS performs the dynamic clustering accordingly. For example, when $p$ is small, the PS assumes that each worker will continue with the same straggling behavior from the past iteration.

\begin{figure}[t]
	\centering  \includegraphics[width=0.5\columnwidth]{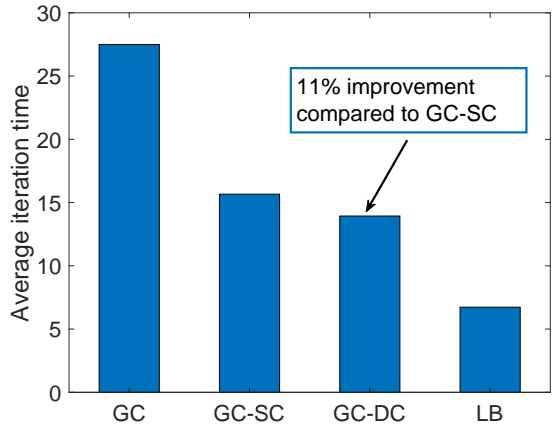}
	\vspace{-1em}
	\caption{Average per-iteration completion time under the Gilbert-Elliot model with homogeneous workers for $K=12$, $P=4$, $r=2$, $n=2$.}
	\label{sim1}
\end{figure}

In the first simulation, we consider the scenario with $K=12$ workers and the dataset is divided into $K = 12$ mini-batches. We set $r = 2$; that is, two partial gradient computations, each corresponding to a different mini-batch, can be computed by each worker at each iteration. We take $P=4$ such that four equal-size clusters are formed. We set $n=2$ and let $6$ of the total $12$ workers start at the slow state, i.e., initially we have $6$ straggling workers. In Fig.~\ref{sim1}, we plot the average per-iteration completion time of the original GC scheme, GC scheme with static clustering (GC-SC), GC scheme with the proposed dynamic clustering (GC-DC), and a lower bound, denoted by LB. Here, the lower bound is obtained by assuming that the full gradient is recovered as soon as the earliest $P\times(\ell-r+1)$ workers finish their computations at each iteration, independently of the codeword assignment matrix. We remark that this lower bound is rather an idealistic scenario as it requires the perfect knowledge of computation times at each iteration as well as $n=P$, i.e., all workers can be assigned to all the clusters.

We observe in Fig.~\ref{sim1} that clustering schemes significantly improve the performance compared to the original GC scheme. The best performance is achieved when the dynamic clustering is implemented, although the performance improvement with respect to GC-SC is smaller than the performance improvement with respect to plain GC by implementing clustering.

\begin{figure}[t]
    \begin{center}
    \subfigure[]{%
    \includegraphics[width=0.49\columnwidth]{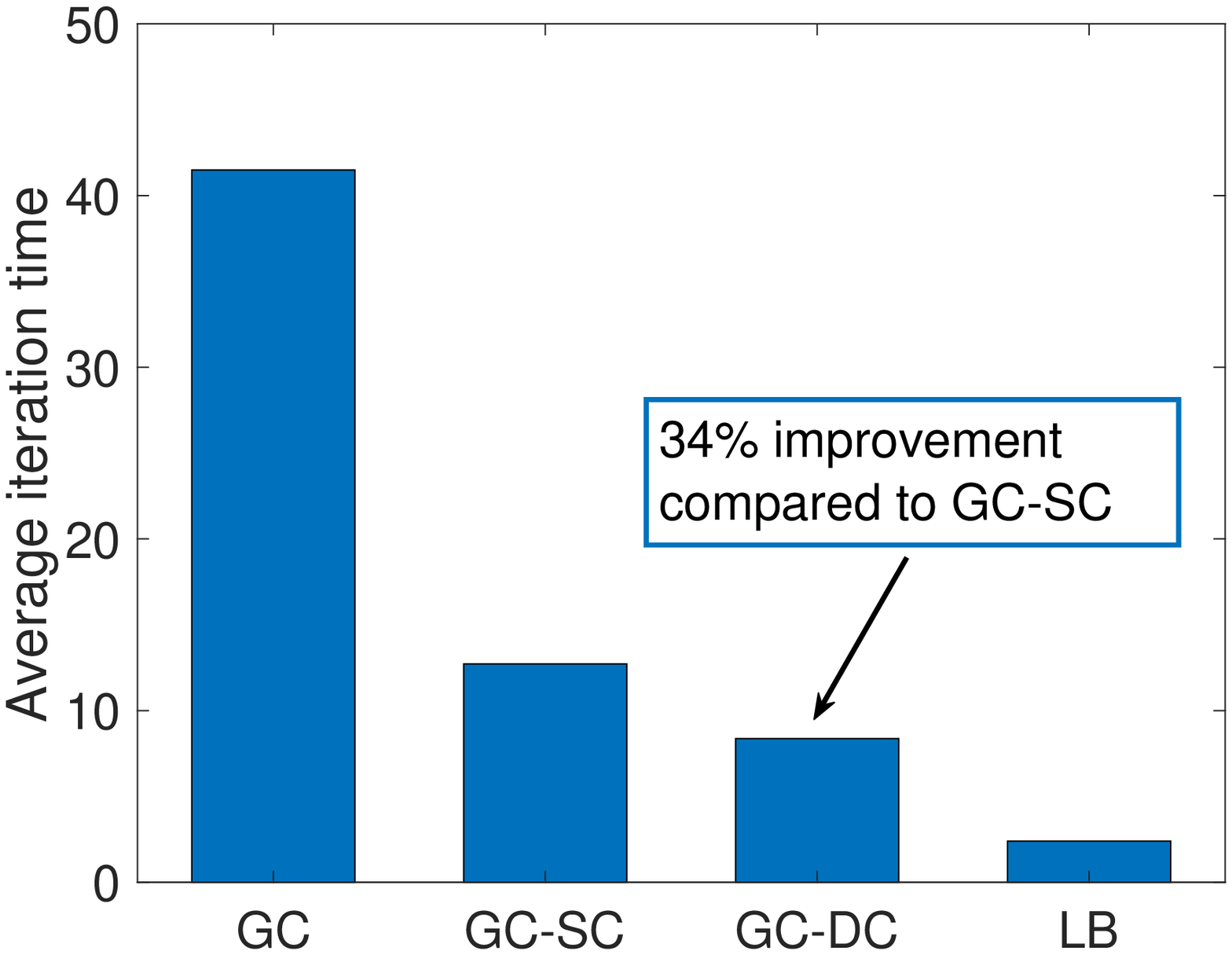}}
    \subfigure[]{%
    \includegraphics[width=0.49\columnwidth]{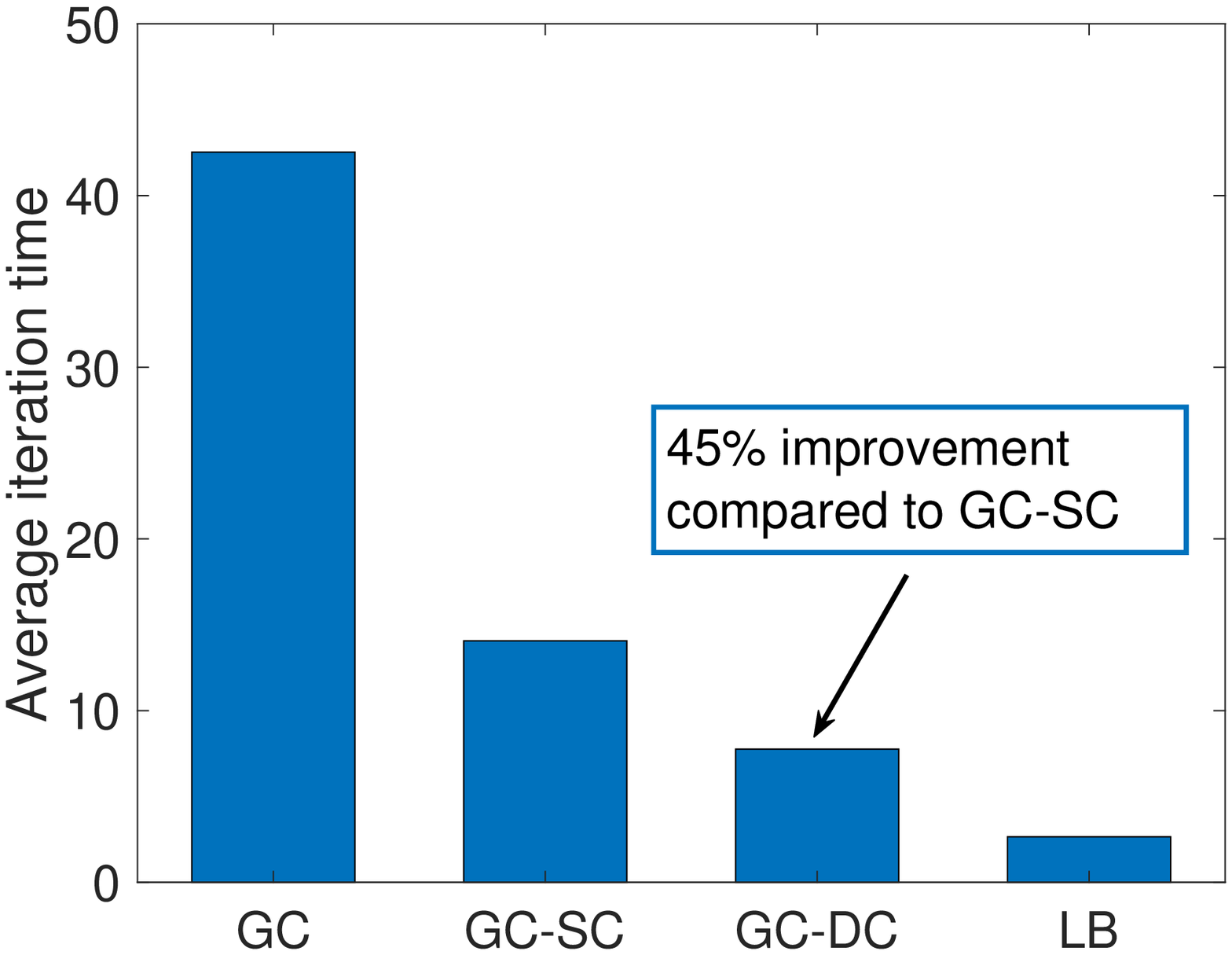}}\\
    \end{center}
    \caption{Average per-iteration completion time under the Gilbert-Elliot model with homogeneous workers for $K=20$, $P=5$, $r=3$, $n=3$ (a) under imperfect SSI, (b) under perfect SSI.}
    \label{sim2and3}        	
\end{figure}

In the second simulation, we set $K=20$, $P=5$, $r=3$, and $n=3$. We start with $10$ stragglers initially. In this case, we observe in Fig.~\ref{sim2and3}(a) that the GC-DC scheme still performs the best and this time the performance improvement compared to the GC-SC scheme (approximately $34\%$) is much more significant. This is due to the increase in the cluster size $\ell$ and the number of assigned clusters $n$, which together increase the dynamic clustering capability of the proposed greedy algorithm.

In the above simulations, we have considered the case in which the PS does not know the exact straggler realization at the beginning of an iteration, and uses previous observation to implement the dynamic clustering strategy. In the third simulation in Fig.~\ref{sim2and3}(b), we consider the same scenario as in the second simulation, but assume that the PS knows the exact straggler realization at the beginning of each iteration, which we call perfect straggler state information (SSI). That is, in the case of perfect SSI, the PS knows exactly which workers will straggle in the current iteration, and therefore, the proposed dynamic clustering algorithm does not suffer from transitions in the straggling behavior from one iteration to the next. In this case we see similar trends as in Fig.~\ref{sim2and3}(a), but observe that the GC-DC scheme results in a larger improvement in the average per-iteration completion time (around $45\%$) than that of the imperfect SSI case.

\subsection{Heterogeneous Worker Model}\label{num_res_het}
In this model, we assume that workers have different computation rates $\mu_k$, $k \in [K].$ In this case, we specify a straggling threshold $\tau>0$, and a worker $k$ is treated as a straggler if $\mu_k < \tau$. 

\subsubsection{Gilbert-Elliot Model with Heterogeneous Workers}

First, we consider a similar model as in Section~\ref{homogenous_workers} and consider the case in which each worker's straggling behavior is modeled by a two-state Markov chain such that $\mu_k = \mu_{k,f}$ if worker $k$ is not straggling and $\mu_k = \mu_{k,s}$ if worker $k$ is a straggler. At the beginning of each iteration, a worker's straggling mode switches with probability $p$. In this case, first we sample the non-straggling computation rates of each worker $\mu_{k,f}$ uniformly at random from the interval $[0,5]$ and set $\alpha_k=0.01$, $p=0.05$ for $k \in [K]$. We model the straggling computation rates of workers $\mu_{k,s}$ such that for worker $k$ we have $\mu_{k,s} = \frac{\mu_{k,f}}{10}$, $k\in [K]$. That is, in the straggling mode, each worker is $10\times$ slower than its typical non-straggling performance, which is motivated by the measurements taken over Amazon EC2 clusters that indicate a similar performance drop in the straggling mode \cite{Yang19}. With this, computation rates of the workers in the straggling mode are uniformly distributed in $[0,0.5]$. We assume that the non-straggling computation rates $\mu_{k,f}$ are known to the PS for $k\in[K]$ after a certain number of iterations and from these, the PS can deduce the straggling computation rates $\mu_{k,s}$. 

Equipped with these, after each iteration, the PS is informed about the straggling status of each worker and performs the proposed greedy dynamic clustering scheme with a modification as follows: Instead of ordering the workers according to (\ref{order}), we order them according to their rates $\mu_k$, $k\in[K]$. In this case, once its turn comes, each cluster selects the fastest available worker first rather than selecting the one with the smallest index first.

We note that since the computation rates are sampled randomly, a worker's straggling computation rate can still be higher than another worker's non-straggling rate. To account for these scenarios, we set the straggling threshold $\tau=0.5$. That is, as long as a worker's rate is below $0.5$ we treat that worker as a straggler. We did not utilize such a threshold in the homogeneous worker model since in that case workers have identical computation rates $\mu_f$ and $\mu_s$ in the non-straggling and straggling states, respectively, such that $\mu_s < \mu_f$.

\begin{figure}[t]
    \begin{center}
    \subfigure[]{%
    \includegraphics[width=0.49\columnwidth]{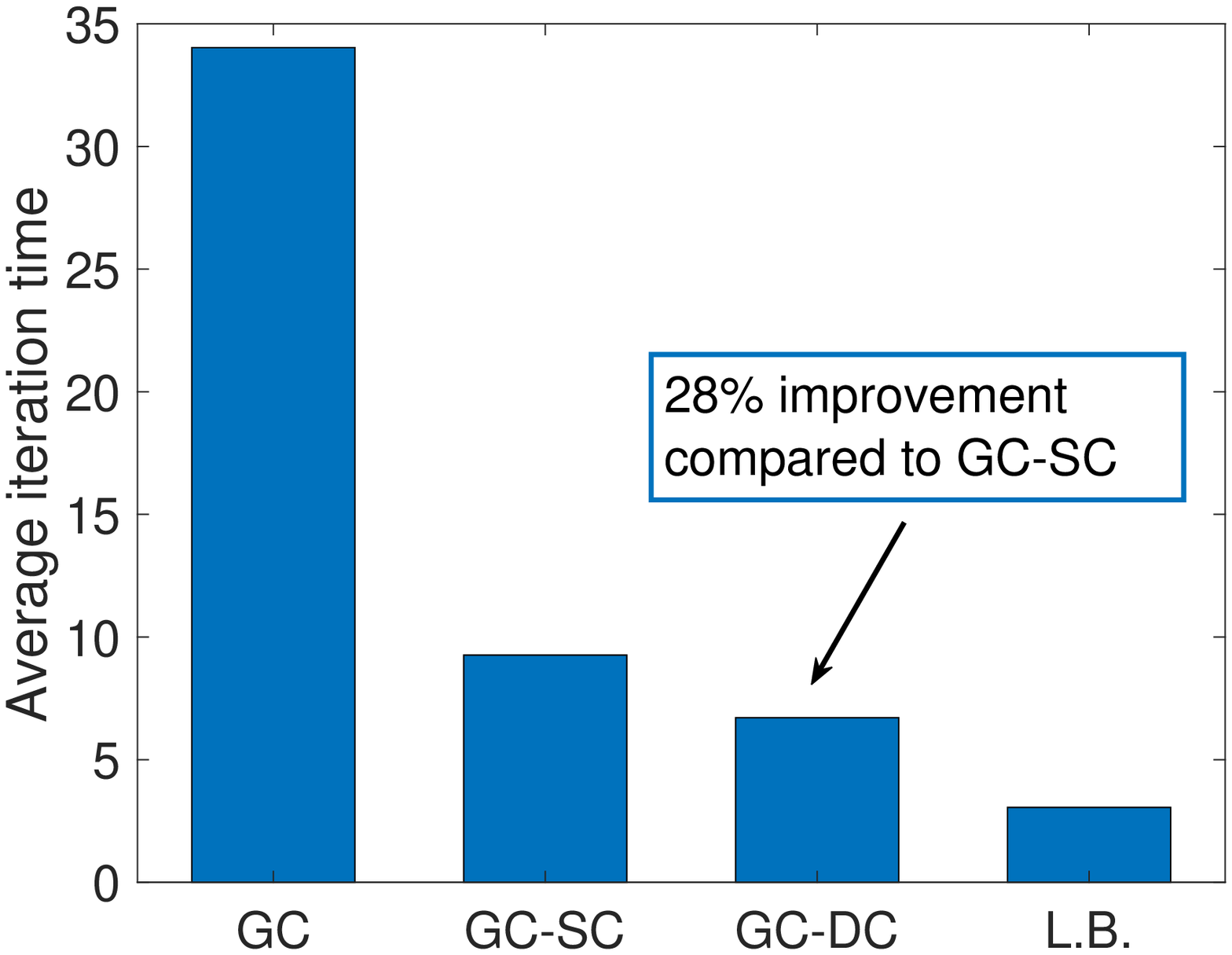}}
    \subfigure[]{%
    \includegraphics[width=0.49\columnwidth]{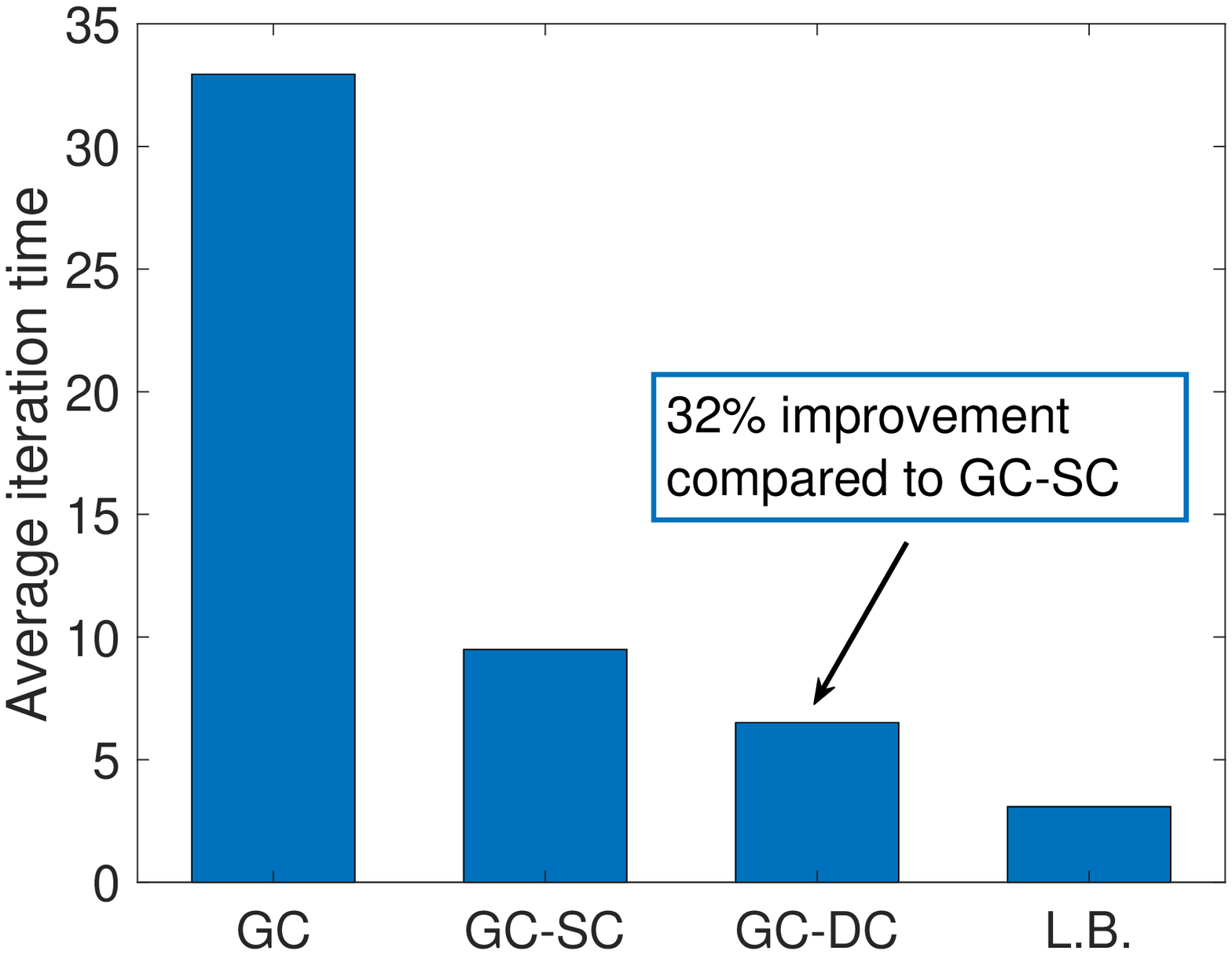}}\\
    \end{center}
    \caption{Average per-iteration completion time under the Gilbert-Elliot model with heterogeneous workers for $K=20$, $P=5$, $r=3$, $n=3$, and $\tau=0.5$ (a) under imperfect SSI, (b) under perfect SSI.}
    \label{hetero_gilber}
\end{figure}

Simulations results for this setup are provided in Fig.~\ref{hetero_gilber}. These results are averaged over $30$ independent simulations for a fixed $\mathbf{A}_{cluster}$ that is generated according to the procedure described in Section~\ref{sect:code_dist}. We observe in Figs~\ref{hetero_gilber}(a) and (b) that the GC-DC scheme outperforms the static clustering schemes, namely GC and GC-SC. As expected, the performance improvement is larger in the case of perfect SSI.

\subsubsection{Heterogeneous Workers with Time-Varying Rates}

So far, we have modeled the straggling behavior based on a Gilbert-Elliot mode. In this subsection, instead of a two-state Markov chain model, we consider that the straggling parameters of the workers are time-varying. We assume that each worker samples its rate uniformly at random from the interval $[0,5]$ and set $\alpha_k=0.01$ for all $k \in [K]$. We assume that at the beginning of each iteration, each worker re-samples its rate with probability $p$ such that with probability $1-p$ its rate stays the same. That is, we have
\begin{align}
    \mu_{k, t+1} = (1-a_{t+1})\mu_{k,t} + a_{t+1}\cdot U[0,5],
\end{align}
where, $\mu_{k,t}$ denotes the rate of worker $k$ at iteration $t$, $a_t$ is an i.i.d.~Bernoulli$(p)$ random variable, i.e., $\mathbb{P}(a_t=1) = p, \forall t$, and $U[a,b]$ denotes a uniform random variable over interval $[a,b]$. In simulations, we use the scenario in the Fig.~\ref{sim2and3} and start with $10$ stragglers. We initialize the rates of stragglers with $\mu_{k,0} = {U}[0, \tau)$ and rates of non-straggling workers with $\mu_{k,0} = U[\tau, 5]$. In this setup, we set $p=0.05$.

Since the computation capabilities of the workers are not identical, we apply the proposed greedy dynamic clustering scheme with the same modification as above. We note that this model requires the workers to accurately detect their computation rates at the end of each iteration and send them to the PS before the next iteration starts.

\begin{figure}[t]
    \begin{center}
    \subfigure[]{%
    \includegraphics[width=0.49\columnwidth]{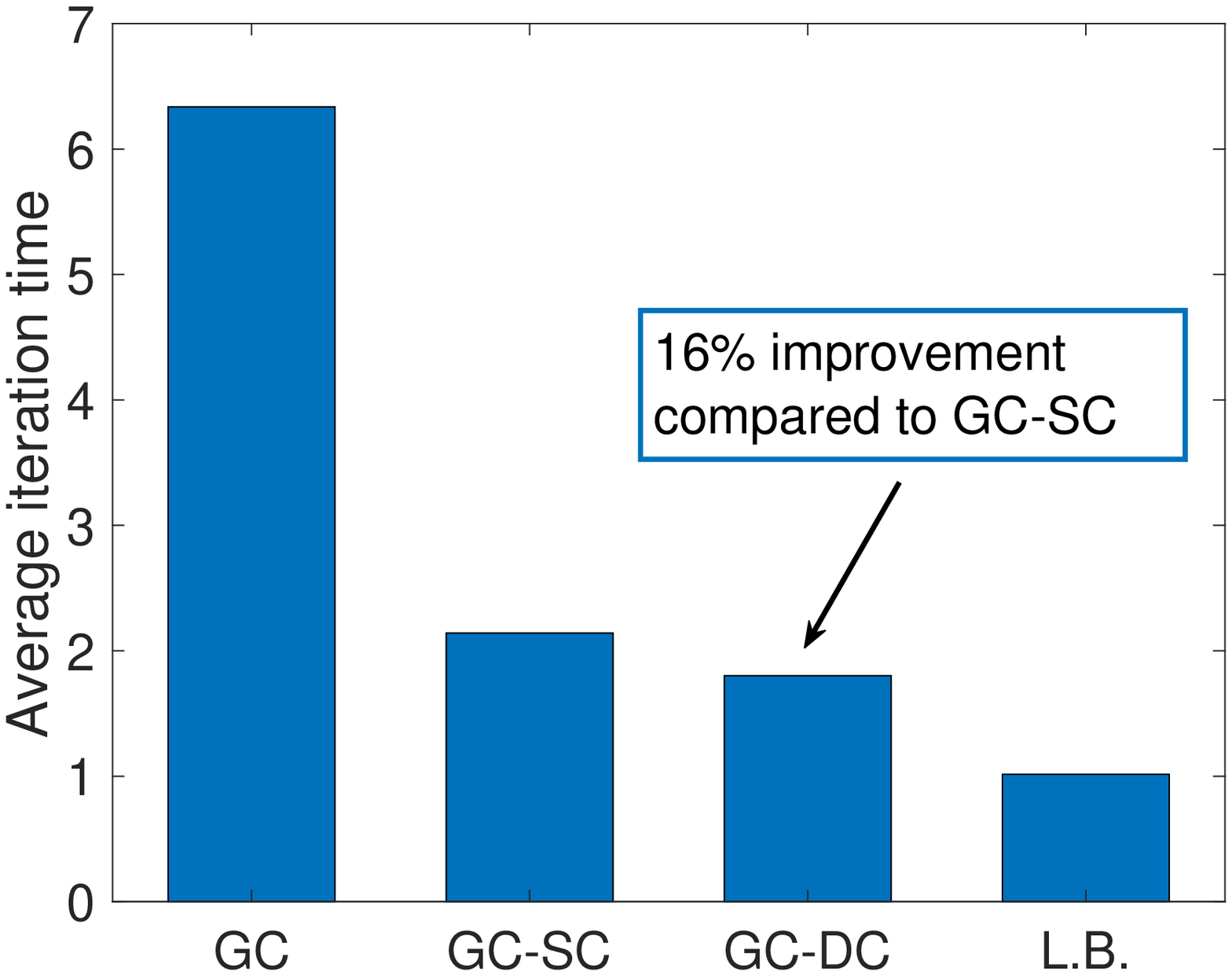}}
    \subfigure[]{%
    \includegraphics[width=0.49\columnwidth]{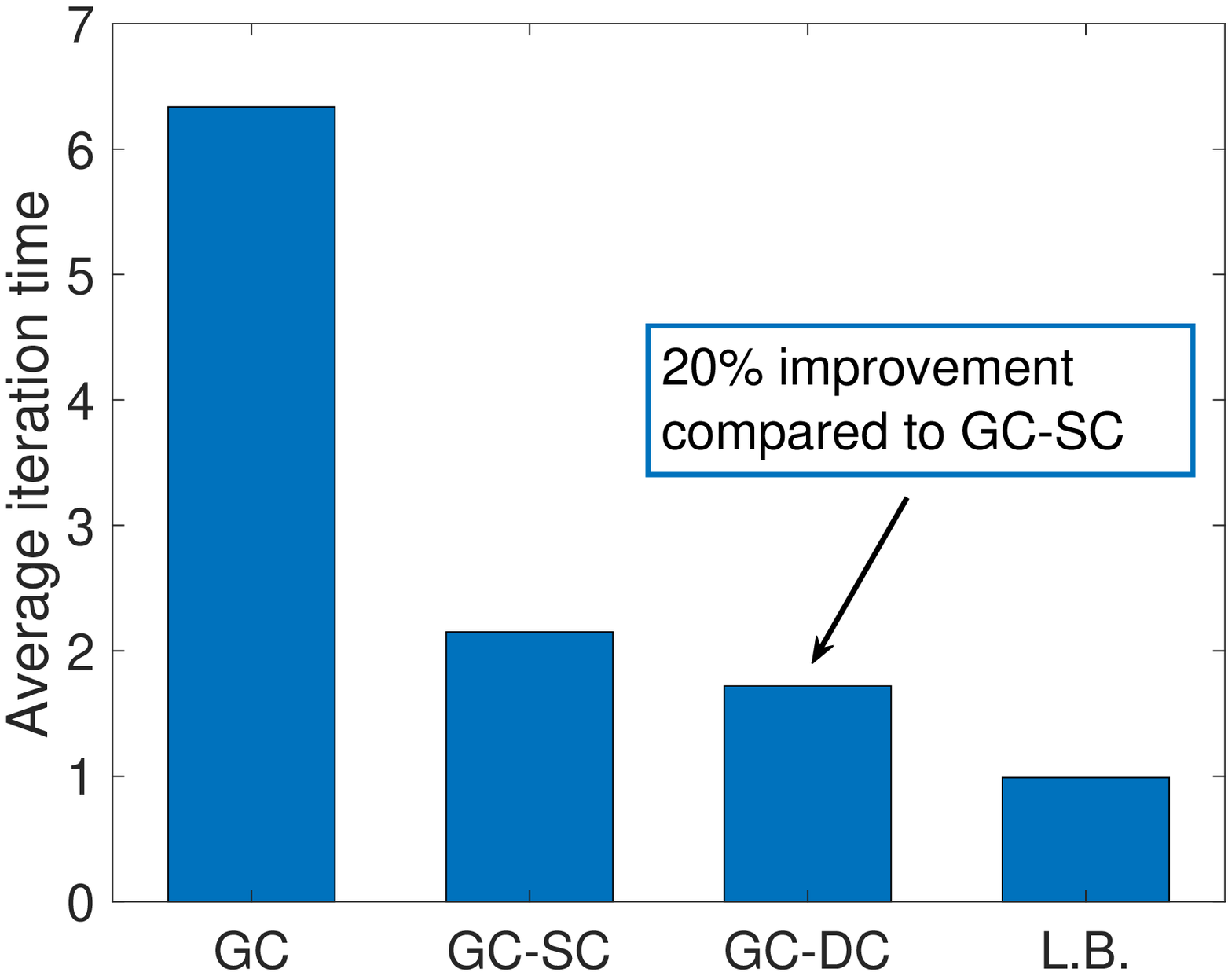}}\\
    \end{center}
    \caption{Average per-iteration completion time under the heterogeneous worker model with time-varying rates for $K=20$, $P=5$, $r=3$, $n=3$, and $\tau=1$ (a) under imperfect SSI, (b) under perfect SSI.}
    \label{sim2and3_thres1}
\end{figure}

First, we consider the case in which $\tau=1$. In this case, we observe in Figs.~\ref{sim2and3_thres1}(a) and (b) that the GC-DC scheme outperforms the GC and GC-SC schemes but the improvement compared to the GC-SC scheme is not significant. In fact, we see that in the case of perfect SSI the improvement is around $20\%$ compared to the GC-SC scheme whereas when the straggler realizations are not known to the PS in advance this improvement drops to approximately $16\%$. 

\begin{figure}[t]
    \begin{center}
    \subfigure[]{%
    \includegraphics[width=0.49\columnwidth]{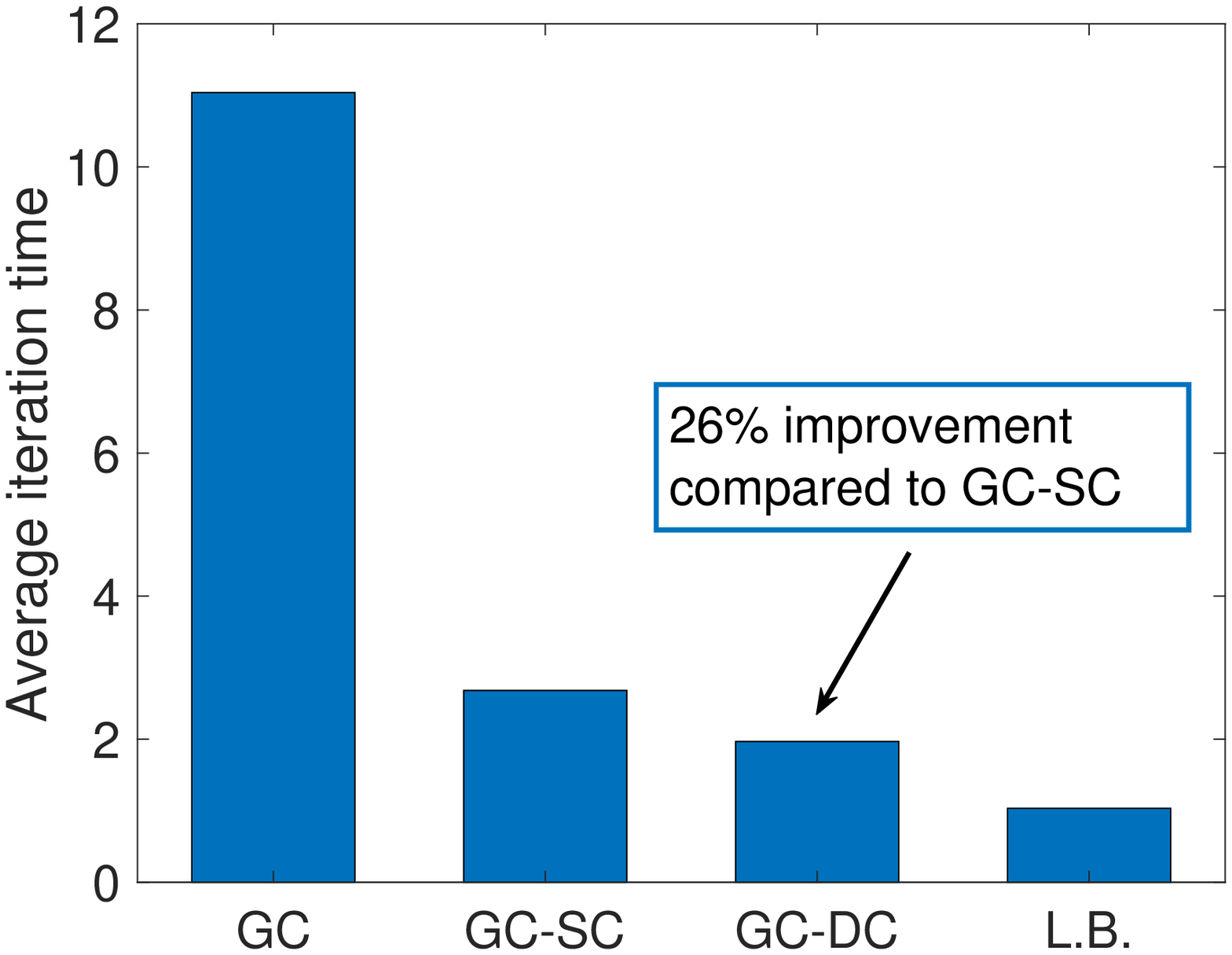}}
    \subfigure[]{%
    \includegraphics[width=0.49\columnwidth]{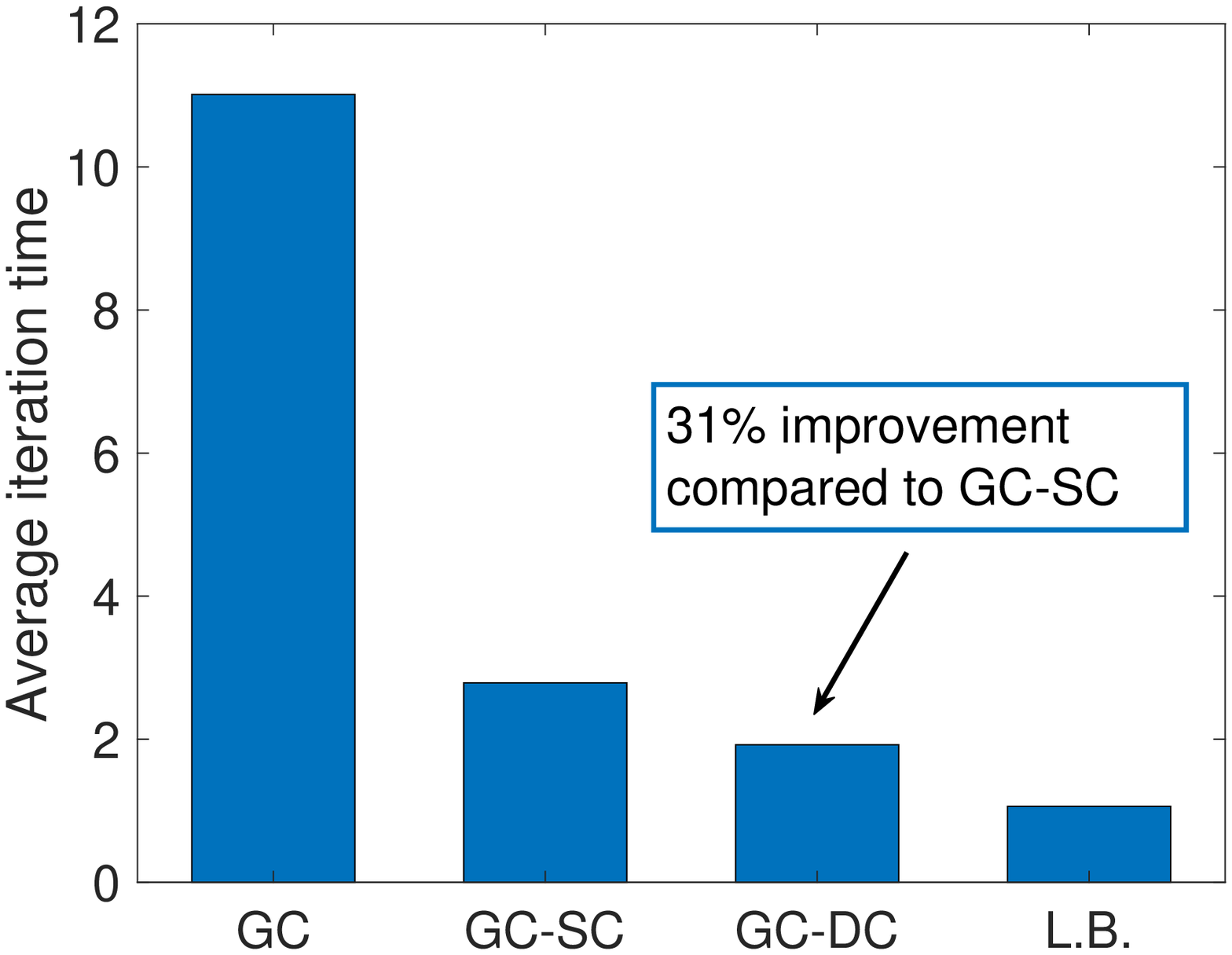}}\\
    \end{center}
    \caption{Average per-iteration completion time under the heterogeneous worker model with time-varying rates for $K=20$, $P=5$, $r=3$, $n=3$, and $\tau=0.1$ (a) under imperfect SSI, (b) under perfect SSI.}
    \label{sim2and3_thres_point1}
\end{figure}

Next, we set $\tau=0.1$ such that the proposed greedy dynamic clustering scheme specifically targets the slowest workers and carefully places them across clusters. In Figs.~\ref{sim2and3_thres_point1}(a) and (b), we observe that the GC-DC scheme performs the best and the improvement compared to the GC-SC scheme is more significant. We also note that in Fig.~\ref{sim2and3_thres_point1}, the performance improvement is larger but the average iteration times are also larger for all three schemes compared to the case in Fig.~\ref{sim2and3_thres1}. This is because when $\tau=0.1$, we initialize the rates of the workers considering $10\times$ slower stragglers compared to when $\tau=1$. We finally note that all the simulation results given in Figs.~\ref{sim2and3_thres1} and~\ref{sim2and3_thres_point1} are averaged over 30 independent simulations for a fixed $\mathbf{A}_{cluster}$ that is generated according to the procedure described in Section~\ref{sect:code_dist}.

\section{Discussion \& Conclusions}\label{conc}

In this work, we considered coded computing for large-scale distributed learning problems in the presence of straggling workers, and introduced a novel scheme, called GC-DC, to reduce the average per-iteration completion time of the static GC schemes. GC-DC employs the GC scheme with clustering introduced in \cite{Ozfatura19}, and assigns additional data to the workers without increasing the per-iteration computation load at each worker compared to the original GC scheme. By utilizing the extra degree-of-freedom offered by additional data, but without increasing the computation load at each iteration, the proposed GC-DC scheme dynamically assigns workers to different clusters at each iteration, in order to distribute the stragglers to clusters as uniformly as possible. Under a time-correlated straggler model, GC-DC can improve the overall computation speed by dynamically adapting to the straggling behavior. We showed through numerical simulations, for both homogeneous and heterogeneous worker models, that the proposed GC-DC scheme can drastically improve the average per-iteration completion time without an increase in the communication load.

We would like to highlight that the proposed redundant data assignment approach with dynamic computations is a fairly general paradigm, and the proposed cluster-based GC approach is only one of many possible coding techniques that can be employed. A possible future research direction is considering heterogeneous cluster sizes. The proposed model assumes that number of workers in each cluster is fixed. That is, cluster sizes are equal to $\ell = \frac{K}{P}$. One can consider varying the cluster sizes to further decrease the average iteration time. Also, in the proposed technique, workers are assigned to clusters based on an order that does not change during the assignment process. To improve the performance, one can consider adaptively changing this worker assignment order. Another potential research direction is to consider a more complex straggling behaviour across the workers, such as non-Markovian, Markovian with higher memory, or correlated straggling behaviour across workers. Such models would require considering all the past straggling behavior when making dynamic clustering assignments, and reinforcement learning techniques can be employed to find the policy that chooses the best code or best clustering strategy to be used at each iteration. 



\bibliographystyle{unsrt}
\bibliography{IEEEabrv,ref,lib_v5}

\begin{thebibliography}{10}

\bibitem{Buyukates21a}
B.~Buyukates, E.~Ozfatura, S.~Ulukus, and D.~Gunduz.
\newblock Gradient coding with dynamic clustering for straggler mitigation.
\newblock In {\em IEEE ICC}, June 2021.

\bibitem{CC.1}
K.~Lee, M.~Lam, R.~Pedarsani, D.~Papailiopoulos, and K.~Ramchandran.
\newblock Speeding up distributed machine learning using codes.
\newblock {\em IEEE Transactions on Information Theory}, 64(3):1514--1529,
  March 2018.

\bibitem{CC.2}
N.~Ferdinand and S.~C. Draper.
\newblock Hierarchical coded computation.
\newblock In {\em 2018 IEEE International Symposium on Information Theory
  (ISIT)}, pages 1620--1624, June 2018.

\bibitem{CC.3}
R.~K. Maity, A.~S. Rawat, and A.~Mazumdar.
\newblock Robust gradient descent via moment encoding with {LDPC} codes.
\newblock {\em {SysML} Conference}, February 2018.

\bibitem{CC.4}
S.~Li, S.~M.~M. Kalan, Q.~Yu, M.~Soltanolkotabi, and A.~S. Avestimehr.
\newblock Polynomially coded regression: Optimal straggler mitigation via data
  encoding.
\newblock May 2018.
\newblock Available on arXiv:1805.09934.

\bibitem{CC.6}
M.~Fahim, H.~Jeong, F.~Haddadpour, S.~Dutta, V.~Cadambe, and P.~Grover.
\newblock On the optimal recovery threshold of coded matrix multiplication.
\newblock In {\em Allerton Conference}, October 2017.

\bibitem{CC.7}
Q.~Yu, M.~Maddah-Ali, and S.~Avestimehr.
\newblock Polynomial codes: an optimal design for high-dimensional coded matrix
  multiplication.
\newblock In {\em NIPS}, December 2017.

\bibitem{Yu18}
Q.~Yu, M.~A. Maddah-Ali, and A.~S. Avestimehr.
\newblock Straggler mitigation in distributed matrix multiplication:
  {Fundamental} limits and optimal coding.
\newblock In {\em IEEE ISIT}, June 2018.

\bibitem{Dutta18c}
S.~Dutta, Z.~Bai, H.~Jeong, T.~M. Low, and P.~Grover.
\newblock A unified coded deep neural network training strategy based on
  generalized polydot codes.
\newblock In {\em IEEE ISIT}, June 2018.

\bibitem{CC.8}
H.~Park, K.~Lee, J.~Sohn, C.~Suh, and J.~Moon.
\newblock Hierarchical coding for distributed computing.
\newblock In {\em IEEE ISIT}, June 2018.

\bibitem{CC.11}
S.~Kiani, N.~Ferdinand, and S.~C. Draper.
\newblock Exploitation of stragglers in coded computation.
\newblock In {\em IEEE ISIT}, June 2018.

\bibitem{CC.13}
A.~B. Das, L.~Tang, and A.~Ramamoorthy.
\newblock {$C^3LES$}: Codes for coded computation that leverage stragglers.
\newblock In {\em IEEE ITW}, November 2018.

\bibitem{Ozfatura19b}
E.~Ozfatura, S.~Ulukus, and D.~Gunduz.
\newblock Distributed gradient descent with coded partial gradient
  computations.
\newblock In {\em IEEE ICASSP}, May 2019.

\bibitem{Mallick19}
A.~Mallick, M.~Chaudhari, and G.~Joshi.
\newblock Fast and efficient distributed matrix-vector multiplication using
  rateless fountain codes.
\newblock In {\em IEEE ICASSP}, May 2019.

\bibitem{Ozfatura18}
E.~Ozfatura, D.~Gunduz, and S.~Ulukus.
\newblock Speeding up distributed gradient descent by utilizing non-persistent
  stragglers.
\newblock In {\em IEEE ISIT}, July 2019.

\bibitem{Yang19}
C.~S. Yang, R.~Pedarsani, and A.~S. Avestimehr.
\newblock Timely coded computing.
\newblock In {\em IEEE ISIT}, July 2019.

\bibitem{Yang19c}
Y.~Yang, M.~Interlandi, P.~Grover, S.~Kar, S.~Amizadeh, and M.~Weimer.
\newblock Coded elastic computing.
\newblock In {\em IEEE ISIT}, July 2019.

\bibitem{Park19}
H.~Park and J.~Moon.
\newblock Irregular product coded computation for high-dimensional matrix
  multiplication.
\newblock In {\em IEEE ISIT}, July 2019.

\bibitem{Bitar19}
R.~Bitar, Y.~Xing, Y.~Keshtkarjahromi, V.~Dasari, S.~E. Rouayheb, and
  H.~Seferoglu.
\newblock Private and rateless adaptive coded matrix-vector multiplication.
\newblock September 2019.
\newblock Available on arXiv: 1909.12611.

\bibitem{Sun19d}
Y.~Sun, J.~Zhao, and D.~Gunduz.
\newblock Heterogeneous coded computation across heterogeneous workers.
\newblock In {\em IEEE Globecom}, December 2019.

\bibitem{Hasircioglu20}
B.~Hasircioglu, J.~Gomez-Vilardebo, and D.~Gunduz.
\newblock Bivariate polynomial coding for exploiting stragglers in
  heterogeneous coded computing systems.
\newblock January 2020.
\newblock Available on arXiv: 2001.07227.

\bibitem{Buyukates19c}
B.~{Buyukates} and S.~{Ulukus}.
\newblock Timely distributed computation with stragglers.
\newblock {\em IEEE Transactions on Communications}, 68(9):5273--5282,
  September 2020.

\bibitem{Ozfatura20}
E.~Ozfatura, B.~Buyukates, D.~Gunduz, and S.~Ulukus.
\newblock Age-based coded computation for bias reduction in distributed
  learning.
\newblock In {\em IEEE Globecom}, December 2020.

\bibitem{UCCT.1}
R.~Tandon, Q.~Lei, A.~G. Dimakis, and N.~Karampatziakis.
\newblock Gradient coding: Avoiding stragglers in distributed learning.
\newblock In {\em ICML}, August 2017.

\bibitem{UCCT.2}
M.~Ye and E.~Abbe.
\newblock Communication-computation efficient gradient coding.
\newblock In {\em ICML}, July 2018.

\bibitem{UCCT.3}
W.~Halbawi, N.~Azizan, F.~Salehi, and B.~Hassibi.
\newblock Improving distributed gradient descent using {Reed-Solomon} codes.
\newblock In {\em IEEE ISIT}, June 2018.

\bibitem{Zhang19}
J.~Zhang and O.~Simeone.
\newblock {LAGC: L}azily aggregated gradient coding for straggler-tolerant and
  communication-efficient distributed learning.
\newblock May 2019.
\newblock Available on arXiv: 1905.09148.

\bibitem{Kadhe19}
S.~Kadhe, O.~O. Koyluoglu, and K.~Ramchandran.
\newblock Gradient coding based on block designs for mitigating adversarial
  stragglers.
\newblock In {\em IEEE ISIT}, July 2019.

\bibitem{Ozfatura19}
E.~Ozfatura, D.~Gunduz, and S.~Ulukus.
\newblock Gradient coding with clustering and multi-message communication.
\newblock In {\em IEEE Data Science Workshop}, June 2019.

\bibitem{Li19}
H.~Wang, S.~Guo, B.~Tang, R.~Li, and C.~Li.
\newblock Heterogeneity-aware gradient coding for straggler tolerance.
\newblock In {\em IEEE ICDCS}, July 2019.

\bibitem{Tauz19}
L.~Tauz and L.~Dolecek.
\newblock Multi-message gradient coding for utilizing non-persistent
  stragglers.
\newblock In {\em Asilomar Conference}, November 2019.

\bibitem{Bitar20}
R.~Bitar, M.~Wootters, and S.~E. Rouayheb.
\newblock Stochastic gradient coding for straggler mitigation in distributed
  learning.
\newblock {\em IEEE Journal on Selected Areas in Information Theory},
  1(1):277--291, May 2020.

\bibitem{Charalambides20}
N.~Charalambides, M.~Pilanci, and A.~O. Hero.
\newblock Weighted gradient coding with leverage score sampling.
\newblock In {\em IEEE ICASSP}, May 2020.

\bibitem{Raviv20}
N.~Raviv, I.~Tamo, R.~Tandon, and A.~G. Dimakis.
\newblock Gradient coding from cyclic {MDS} codes and expander graphs.
\newblock {\em IEEE Transactions on Information Theory}, 66(12):7475--7489,
  December 2020.

\bibitem{UCUT.5}
J.~Chen, R.~Monga, S.~Bengio, and R.~Jozefowicz.
\newblock Revisiting distributed synchronous {SGD}.
\newblock April 2016.
\newblock Available on arXiv:1604.00981.

\bibitem{UCUT.2}
S.~Li, S.~M.~M. Kalan, A.~S. Avestimehr, and M.~Soltanolkotabi.
\newblock Near-optimal straggler mitigation for distributed gradient methods.
\newblock In {\em IEEE IPDPS}, May 2018.

\bibitem{UCUT.1}
S.~Dutta, G.~Joshi, S.~Ghosh, P.~Dube, and P.~Nagpurkar.
\newblock Slow and stale gradients can win the race: Error-runtime trade-offs
  in distributed {SGD}.
\newblock In {\em International Conference on Artificial Intelligence and
  Statistics (AISTATS)}, April 2018.

\bibitem{Ferdinand18b}
N.~Ferdinand and S.~C. Draper.
\newblock Anytime stochastic gradient descent: A time to hear from all the
  workers.
\newblock In {\em Allerton Conference}, October 2018.

\bibitem{BehrouziFar18}
A.~Behrouzi-Far and E.~Soljanin.
\newblock On the effect of task-to-worker assignment in distributed computing
  systems with stragglers.
\newblock In {\em Allerton Conference}, October 2018.

\bibitem{UCUT.4}
M.~Mohammadi Amiri and D.~Gunduz.
\newblock Computation scheduling for distributed machine learning with
  straggling workers.
\newblock {\em IEEE Transactions on Signal Processing}, 67(24):6270--6284,
  December 2019.

\bibitem{Korte08}
B~Korte and J.~Vygen.
\newblock {\em Combinatorial Optimization: Theory and Algorithms}.
\newblock Springer, 2008.

\bibitem{entropy}
E.~Ozfatura, S.~Ulukus, and D.~Gunduz.
\newblock Straggler-aware distributed learning: Communication computation
  latency trade-off.
\newblock {\em Entropy, Special Issue on the Interplay Between Storage,
  Computing, and Communications from an Information-Theoretic Perspective},
  22(5):544, May 2020.

\bibitem{Buyukates20a}
B.~Buyukates and S.~Ulukus.
\newblock Age of information with {Gilbert-Elliot} servers and samplers.
\newblock In {\em CISS}, March 2020.

\end{thebibliography}
\end{document}